\documentclass[usegraphicx,useAMS,usenatbib]{mn2e}

\newcommand{\beq}{\begin{equation}}
\newcommand{\beqa}{\begin{eqnarray}}
\newcommand{\eeq}{\end{equation}}
\newcommand{\eeqa}{\end{eqnarray}}
\newcommand{\etahat}{\hat{\eta}}
\newcommand{\rbeta}{R_{\beta}}
\renewcommand{\vec}[1]{\bmath #1}

\title[Resistive jet simulations]{Resistive jet simulations
extending radially self-similar magnetohydrodynamic models}
\author
[M. \v{C}emelji\'{c} et al.]
{M. \v{C}emelji\'{c}$^1$\thanks{E-Mail: miki@tiara.sinica.edu.tw (MC);
jgracia@cp.dias.ie (JG); vlahakis@phys.uoa.gr (NV); tsingan@phys.uoa.gr (KT)},
J. Gracia$^2$\footnotemark[1], N. Vlahakis$^3$\footnotemark[1]
and K. Tsinganos$^3$\footnotemark[1]\\
$^1$Theoretical Institute for Advanced Research in Astrophysics (TIARA),
Academia Sinica and Institute for Astronomy and Astrophysics,\\
National Tsing Hua University, No. 101, Sec. 2, Kuang Fu Rd., Hsinchu 30013, Taiwan\\
$^2$School of Cosmic Physics, Dublin Institute for Advanced Studies,
31 Fitzwilliam Place, Dublin 4, Ireland\\
$^3$IASA and Section of Astrophysics, Astronomy and Mechanics,
Department of Physics, University of Athens,\\
Panepistemiopolis 15784 Zografos, Athens, Greece}

\begin{document}
\date{Received/Accepted}
\pagerange{\pageref{firstpage}--\pageref{lastpage}} \pubyear{2008}

\maketitle

\label{firstpage}

\begin{abstract}
Numerical simulations with self-similar initial and boundary conditions provide a
link between theoretical and numerical investigations of jet dynamics. We
perform axisymmetric resistive magnetohydrodynamic (MHD) simulations for a generalised solution of
the Blandford \& Payne type,
and compare them with the corresponding analytical and numerical ideal-MHD solutions. 
We disentangle the effects of the numerical and physical diffusivity. 
The latter could occur in outflows above an accretion disk, being transferred from the underlying disk
into the disk corona by MHD turbulence (anomalous turbulent diffusivity),
or as a result of ambipolar diffusion in partially ionized flows.
We conclude that while the classical magnetic Reynolds number $R_{\rm m}$ measures the
importance of resistive effects in the induction equation, a new introduced number,
$\rbeta=(\beta/2)R_{\rm m}$ with $\beta$ the plasma beta,
measures the importance of the resistive effects in the energy equation.
Thus, in magnetised jets with $\beta<2$, when $\rbeta \la 1$  resistive effects
are non-negligible and affect mostly the energy equation.
The presented simulations indeed show that for a range of
magnetic diffusivities corresponding to $\rbeta \ga 1$
the flow remains close to the ideal-MHD self-similar solution.
\end{abstract}


\section{Introduction}

Collimated outflows of plasma observed to emerge from the vicinity of a wide spectrum 
of cosmic objects are still a challenge for observational and theoretical astrophysics. 
These outflows play a key role in the transport of angular momentum and energy of 
the accreted gas facilitating thus, for example, star formation. 
Nevertheless, when new observations put more and more severe constraints on models, 
these seem to be still too rudimentary to provide sophisticated answers. 

The starting point of the modeling of jets are the ideal  MHD equations, which can be 
solved analytically by assuming axisymmetry, time-independence and the self-similarity 
ansatz. Analytical models of ideal MHD disk winds (Blandford \& Payne 1982,
re-visited in Vlahakis et al. 2000; hereafter V00), provide not only the first insight 
into the physics of such outflows but equally important they can be used as a test bed of 
more sophisticated  simulations of the resistive MHD system via various numerical codes. 
In Vlahakis \& Tsinganos (1998) general classes of self-consistent ideal-MHD
solutions have been constructed. Two sets of exact MHD outflow models have
been found: meridionally and radially self-similar ones. Previously known studies
were recognised to belong in this more general classification of all available analytical models.
In particular, the V00 study remedied the physically unacceptable feature of the
Blandford \& Payne (1982) terminal 
wind solution which was not causally disconnected from the disk (see also Ferreira \& Casse 
(2004) when a resistive disk is included).

Among the basic problems which still remained to be solved however were the common deficiency that 
all radially self-similar models had, namely, a cut-off of the solution at small cylindrical 
radii and also at some finite height above the disk where they were unphysical. 
The reason for such behaviour is a strong Lorentz force close to the system's axis. The invalid 
analytical solution very close to the axis has been corrected numerically.
Also, a search in the numerical simulations for solutions at larger distances from the
disk has been performed, in Gracia et al. (2006; hereafter GVT06), where ideal-MHD numerical 
simulations with the V00 solution as initial condition have been performed using the
NIRVANA code (version 2.0, Ziegler 1998). These results have been verified also by
using the PLUTO code (Mignone et al. 2007) in the ideal MHD simulations by
Matsakos et al. (2008). 

\begin{figure}
\hspace{1.3cm}\includegraphics[width=6.cm, height=1.cm]{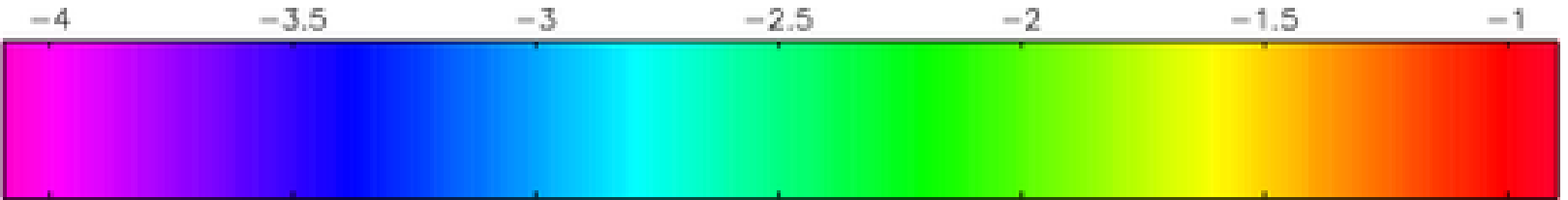}
\includegraphics[width=7.5cm, height=12cm]{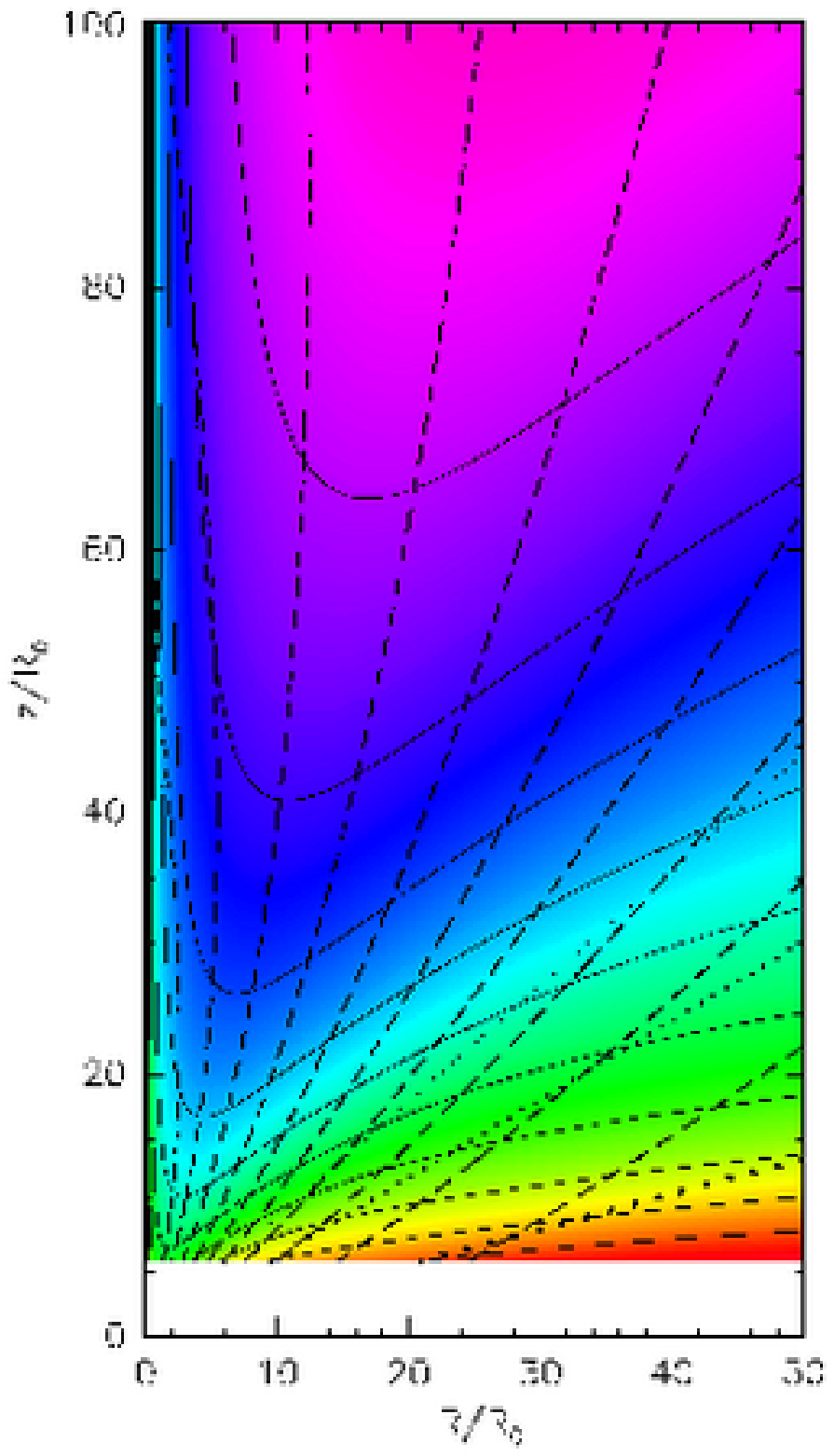}
\caption{Initial conditions for our numerical simulations. The solid
lines represent logarithmically spaced isocontours of density. It is also
shown in colour grading. The dashed lines show poloidal fieldlines or
flowlines. The dotted lines
indicate, from top to bottom, the position of the fast-magnetosonic
({\em small dots}), the Alfv\'en, and the slow-magnetosonic surface ({\em
large dots}), respectively.
}
\label{pocet}
\end{figure}

The next step in the exploitation of the available analytical solutions has been the 
investigation of the dynamical connection of the outflow to the underlying disk 
(see e.g., K\"{o}nigl, 1989; Wardle \& K\"{o}nigl 1993; Li 1995; Ferreira 1997; Casse 
\& Keppens 2004; Zanni et al. 2007), providing some understanding of the formation 
of jets from the accretion disk.
At the same time, since the disk is naturally resistive, the need emerged to go beyond 
the ideal MHD regime. With the magnetic field included, the effects of the
magnetic resistivity, both numerical and physical, had to be addressed, discriminated and analysed.
 
In fully ionized disks an anomalous turbulent diffusivity has to be invoked
to allow for accretion of matter crossing a large-scale magnetic field.
This anomalous turbulent diffusivity may be present in the outflow as well, at least
at distances close to the disk.
In partially ionized disks the physical conductivity is properly described by a tensor,
consisting of three distinct parts corresponding to the ambipolar diffusion, 
the Hall effect, and the Ohmic dissipation (e.g. Wardle \& Ng, 1999; Salmeron et al., 2007).
The dominant mechanism in the outflow above the disk is most likely 
the ambipolar diffusion (e.g., Sano \& Stone 2002; Kunz \& Balbus 2004; Wardle 2007),
and can be appropriately described only by a multifluid MHD.
Nevertheless, in both cases (turbulent or ambipolar diffusion),
a scalar conductivity can capture the basic characteristics of the 
breakdown of ideal MHD related to the magnetic field diffusion and the
resistive heating. 

The effects of the resistive heating in the formation and acceleration of jets from
resistive disks or tori, have been studied in Kuwabara et al. (2000, 2005), concluding
that Joule heating is not playing an essential role in jet formation. However, in these studies
only low values (lower than the critical value that we define below) of resistivity were examined.
Resistive effects have been also studied in Fendt \& \v{C}emelji\'{c} (2002; hereafter FC02).
In this case, however, the energy equation was not solved and
a polytropic equation of state has been assumed instead,
such that the effects of finite resistivity
have not been directly incorporated in the energetics of the problem.
Safier (1993a) working in the ambipolar diffusion regime of outflows
associated with young stars, found that 
the diffusive term in the energy equation cannot be neglected.
The heating of the gas could have significant observational consequences
(see e.g. Safier 1993b; Martin 1996b; Cabrit et al., 1999; Garcia et al., 2001a,b;
O'Brien et al., 2003; Shang et al., 2004).
Safier's (1993a) work also shows why the cooling term in the energy equation 
and the ionization balance equation need to be considered in a full investigation.
He was able to uncover a strong feedback mechanism between the
gas temperature and the ionization fraction (which scales inversely
with the ambipolar diffusion heating rate).
The result of the heating depends also on the geometry of the flow
which strongly affects the adiabatic cooling.
In a spherical outflow this cooling effectively counters the Joule dissipation heating 
(Ruden et al., 1999), while in a disk-driven jet with a small streamline divergence
the adiabatic cooling is relatively unimportant (at least initially).
Interestingly, Joule dissipation can play a role
(although generally not a dominant one) also in magnetically guided
accretion problems (e.g., Martin 1996a).

Numerical resistivity is implicitly present in any numerical simulation, and its various 
effects need to be identified and studied in detail. This is one of the main aims of this paper. 
The other one is to investigate the effects of small physical resistivity, examine the 
stability of the resistive jet solutions and define when the resistivity is "small" and 
when it becomes "large".
The novel approach followed here is that the analytical solution is a stationary reference 
solution, something which was lacking in the previous investigations of resistive MHD flows, e.g.
FC02.

The structure of the paper is as follows. The analytical expressions and
their modification in the setup for numerical simulations are first presented 
in Sec. 2.
Then the resistive-MHD solutions (Sec. 4) are systematically compared to the ideal-MHD ones of Sec. 3.
In Sec. 5 we introduce an extension of the magnetic Reynolds number which quantifies the transition 
from an ideal-like behaviour of the low diffusivity solutions for values of the diffusivity $\eta$ below a 
critical value $\eta_{crit}$, to a transient and erratic behaviour of the solutions for $\eta > \eta_{crit}$.  
The main possible consequences for astrophysical outflows are briefly discussed. A summary of the main results 
is given in the last Sec. 6. 

\section[]{Problem setup}\label{sec2}
\subsection{Governing equations}
The resistive-MHD equations solved by the NIRVANA code are, in SI units:
\beqa
\frac{\partial \rho}{\partial t} + \nabla \cdot (\rho \vec{V})=0 \,,\\
\rho\left[ \frac{\partial\vec{V} }{\partial t}+ \left( \vec{V
}\cdot \nabla\right) \vec{V} \right] + \nabla p +
\rho\nabla \Phi
- \frac{ \nabla \times \vec{B}}{\mu_0} \times \vec{B} = 0 \label{mom2} \,, \\
\frac{\partial\vec{B} }{\partial t}- \nabla \times \left( \vec{V}
\times \vec{B}-\eta \nabla \times \vec{B} \right)= 0 \,, \label{faraday}\\
\rho \left[ {\frac{\partial e}{\partial t}}
+ \left(\vec{V} \cdot \nabla\right)e \right]
+ p(\nabla \cdot\vec{V} )
- \frac{\eta}{\mu_0} \left( \nabla \times \vec{B} \right)^2= 0 \,, \label{enn}\\
 \nabla \cdot \vec{B}=0 \,,
\eeqa
where $\vec{V}$ is the flow velocity, $\vec{B}$ is the magnetic field,
$(\rho, P)$ are the gas density and pressure,
and $\Phi=-{\cal GM}/r$ is the gravitational potential of the central mass
${\cal M}$.
The internal energy (per unit mass) is related to the pressure and density by
\beq
e=\frac{1}{\gamma-1} \ \frac{p}{\rho} \,,
\eeq
where $\gamma$ is the effective polytropic index. The magnetic diffusivity
$\eta$ is assumed constant and
is related to the resistivity $\rho_c=\mu_0\eta$,
where $\mu_0$ is the permeability of vacuum.

\subsection{Initial and boundary conditions}
For the initial and boundary conditions of the simulations
the self-similar solution of V00 is used.
The assumptions of steady-state,
axisymmetry and radial self-similarity result in the following expressions
for the physical quantities in spherical ($r$, $\theta$, $\phi$)
and cylindrical ($Z=r\cos\theta$, $R=r\sin\theta$, $\phi$) coordinates:
\beqa
\frac{\rho}{\rho_0}=\alpha^{x-3/2}\frac{1}{M^2}\,,
\label{rhoss}\\
\frac{p}{p_0}=\alpha^{x-2}\frac{1}{M^{2\gamma}}\,,\\
\frac{\vec{B}_p}{B_0}=-\alpha^{\frac{x}{2}-1}\frac{1}{G^2}\frac{
\sin\theta}{\cos(\psi+\theta)}
\left(\cos\psi \hat{R}+\sin\psi \hat{Z}\right) \,,
\label{bpss}\\
\frac{\vec{V}_p}{V_0}=-\alpha^{-1/4}\frac{M^2}{G^2}\frac{
\sin\theta}{\cos(\psi+\theta)}
\left(\cos\psi \hat{R}+\sin\psi \hat{Z}\right) \,, \\
\frac{{B}_\phi}{B_0}=-\lambda\alpha^{\frac{x}{2}-1}\frac{1-G^2}{G(1-M^2)}\,,\\
\frac{{V}_\phi}{{V}_0}=\lambda\alpha^{-\frac{1}{4}}\frac{G^2-M^2}{G(1-M^2)} \,,
\label{vphiss}
\eeqa
where $\displaystyle \alpha=\frac{R^2}{R_0^2 G^2}$,
$(M\,, G\,, \psi)$ are functions of $\theta$, and
\beq
\displaystyle{
V_0=\frac{1}{\kappa} \sqrt{\frac{ \cal G M }{R_0}} \,, \quad
\rho_0=\frac{B_0^2}{\mu_0 V_0^2} \,, \quad
p_0=\mu\frac{B_0^2}{2\mu_0} \,.
}
\label{norms}
\eeq
Here we decomposed vector quantities in poloidal (index $p$) and
toroidal (index $\phi$) components. We set the solution parameters
to $(x\,, \lambda^2, \mu\,, \kappa\,, \gamma)=
(0.75\,, 136.9\,, 2.99\,, 2\,, 1.05)$, as in the V00 solution.

The diffusivity $\eta$ (which is assumed constant throughout the domain)
is normalised as $\eta = \etahat \ V_0 R_0=\etahat \sqrt{{\cal GM}R_0}/\kappa$, with
$\etahat$ dimensionless.
\begin{figure*}
\includegraphics[width=5.5cm,height=9cm]{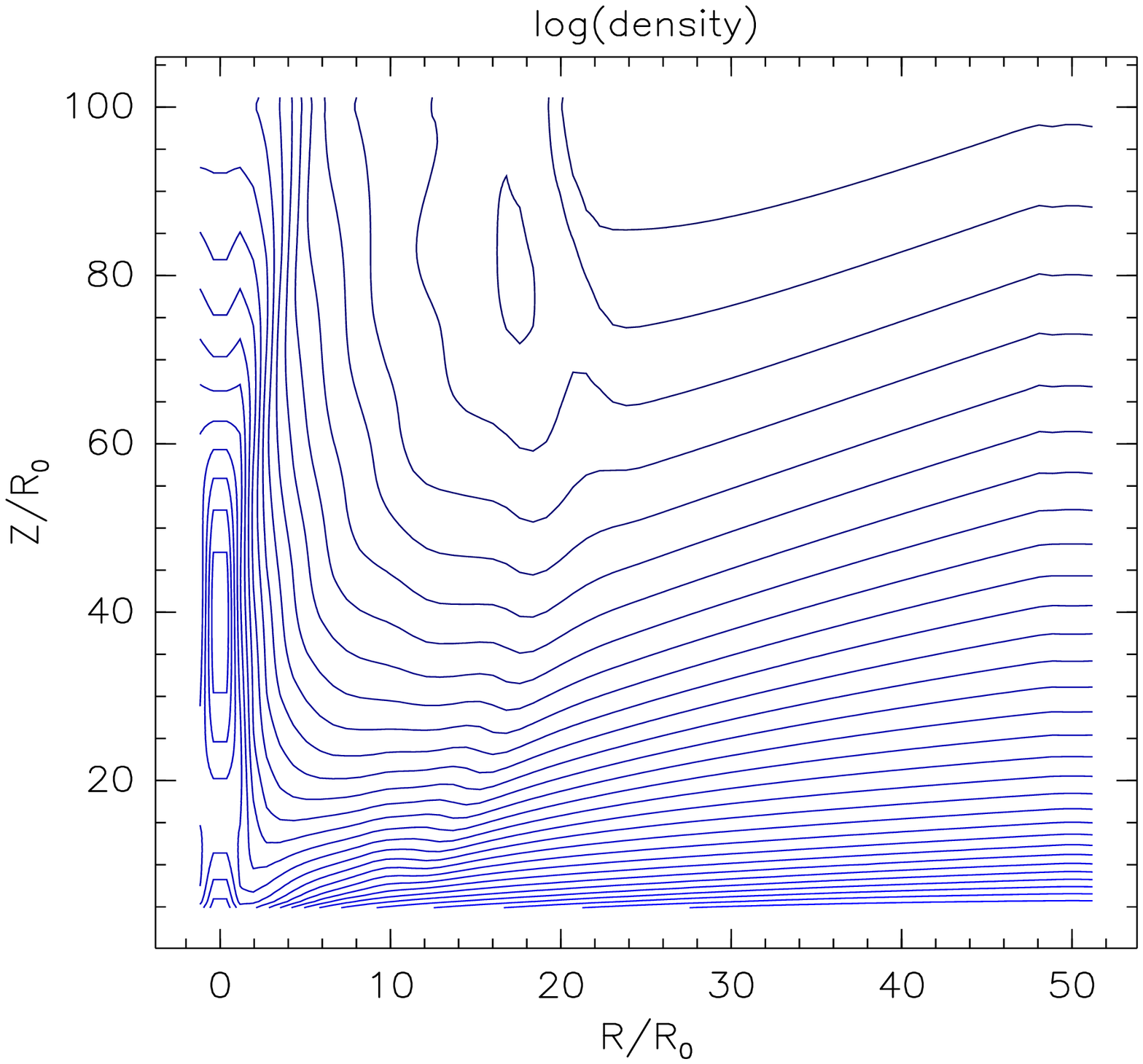}
\includegraphics[width=5.5cm,height=9cm]{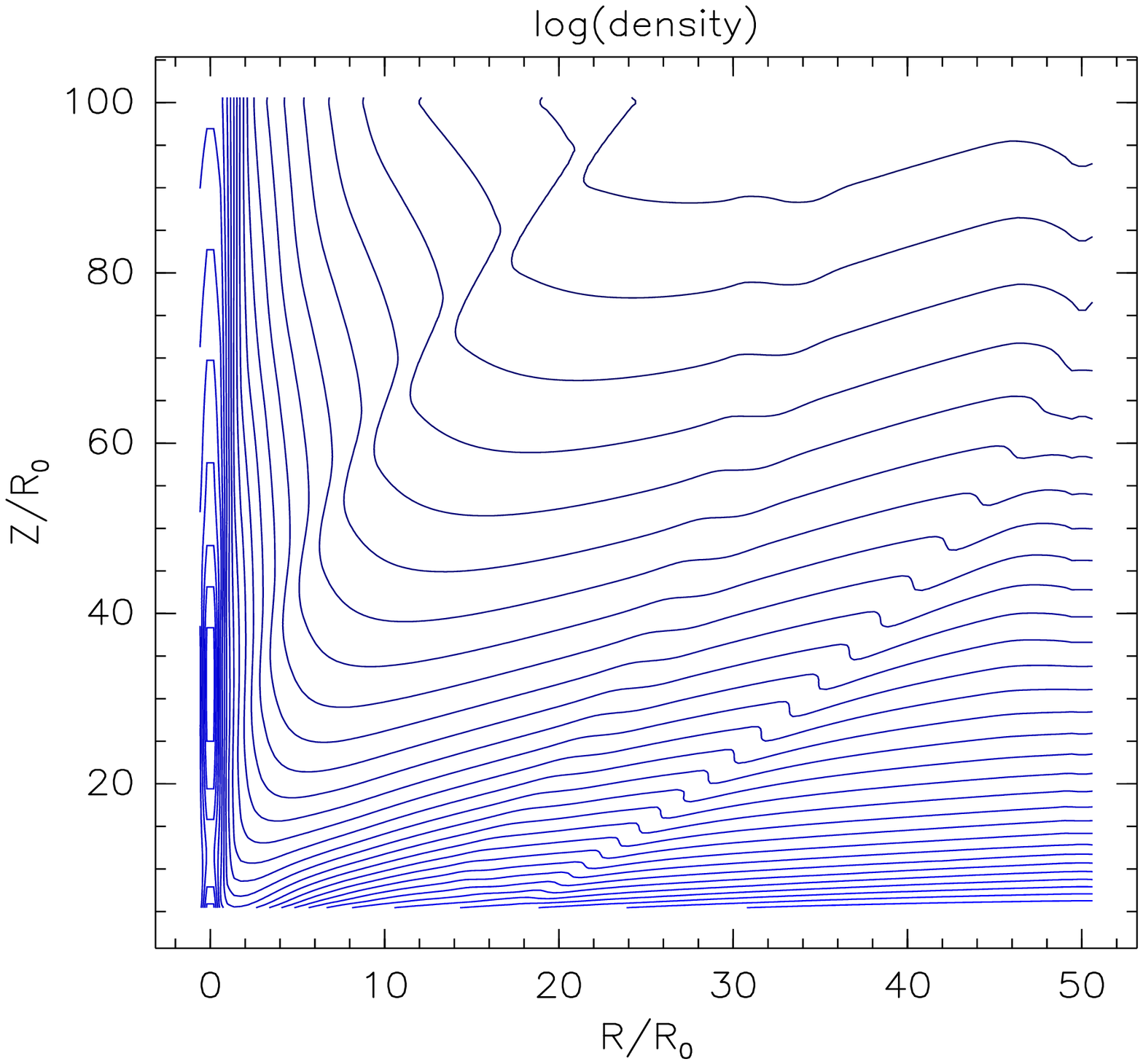}
\includegraphics[width=5.5cm,height=9cm]{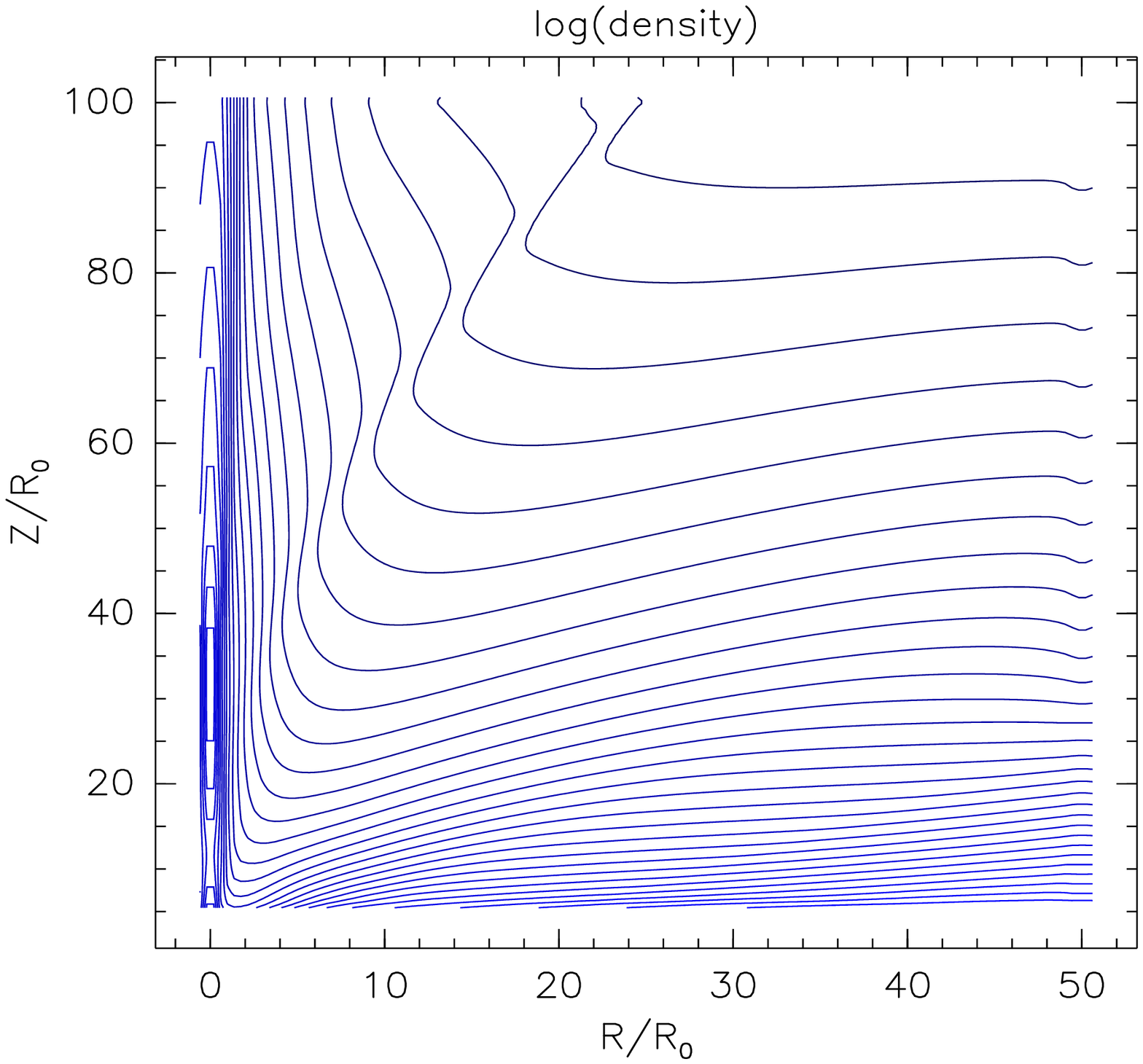}
\caption{Ideal-MHD simulations before, during and after the relaxation,
for the resolution of $R\times Z = (128\times 256)$ grid cells $=([0,50]\times [6,100])R_0$.
Lines denote thirty logarithmically spaced isocontours of density. The times of
simulations are
few thousands, few ten thousands and larger than few hundred thousands of
the Courant time steps, {\em Left} to the {\em Right} panel, respectively.
}
\label{idealdens}
\end{figure*}
The self-similar solution breaks down near the rotation axis.
This becomes evident from the fact that all physical quantities
are proportional to a power of the function $1/\alpha$,
which is divergent on the axis ($R=0$),
see Eqs. (\ref{rhoss}-\ref{vphiss}).
In addition, the analytical solution of V00 is not
provided for $\theta$ smaller than 0.025 rad, measured from the axis.

To perform
numerical simulations in a computational box with the symmetry axis included,
we need to modify/extrapolate the analytical solution. Near the axis, we
extrapolated the missing analytical
solutions for the tabulated functions $G$, $M$ and $\psi$, as described in GVT06.
A similar result can be achieved with less involved extrapolation, as shown in
Matsakos et al. (2008). Modification of the functions G and M means also that
 the pressure/energy is modified near the axis.

For the magnetic field in the vicinity of the symmetry axis there is an
additional problem. With the extrapolated functions $G$ and $\psi$
the magnetic field as given by Eq.~(\ref{bpss}) is not divergence-free.
This leads to the need for suitable modification of the initial magnetic field.
A simple modification is to compute the $B_Z$ component from
the $\hat{Z}$ component of the self-similar expression
\beq
\vec{B}_p=\frac{B_0 R_0^2}{x} \nabla \times
\left( \alpha^{x/2}\frac{\hat{\phi}}{R} \right) \,,
\eeq
and subsequently the radial component $B_R$ by solving the $\nabla\cdot \vec B=0$
with boundary condition $B_R(R=0)=0$.
\begin{figure*}
\includegraphics[width=5.5cm,height=8cm]{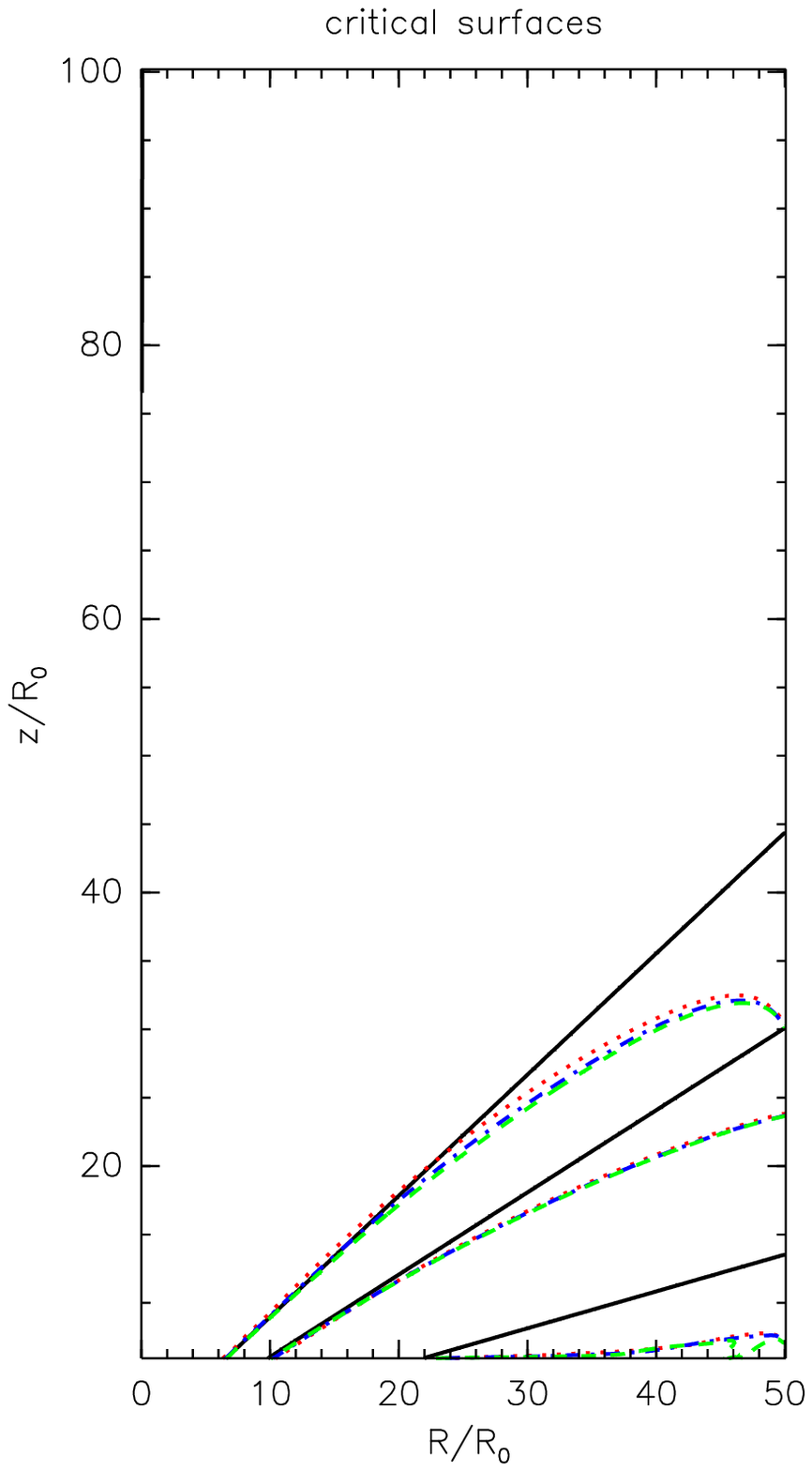}
\includegraphics[width=5.5cm,height=8cm]{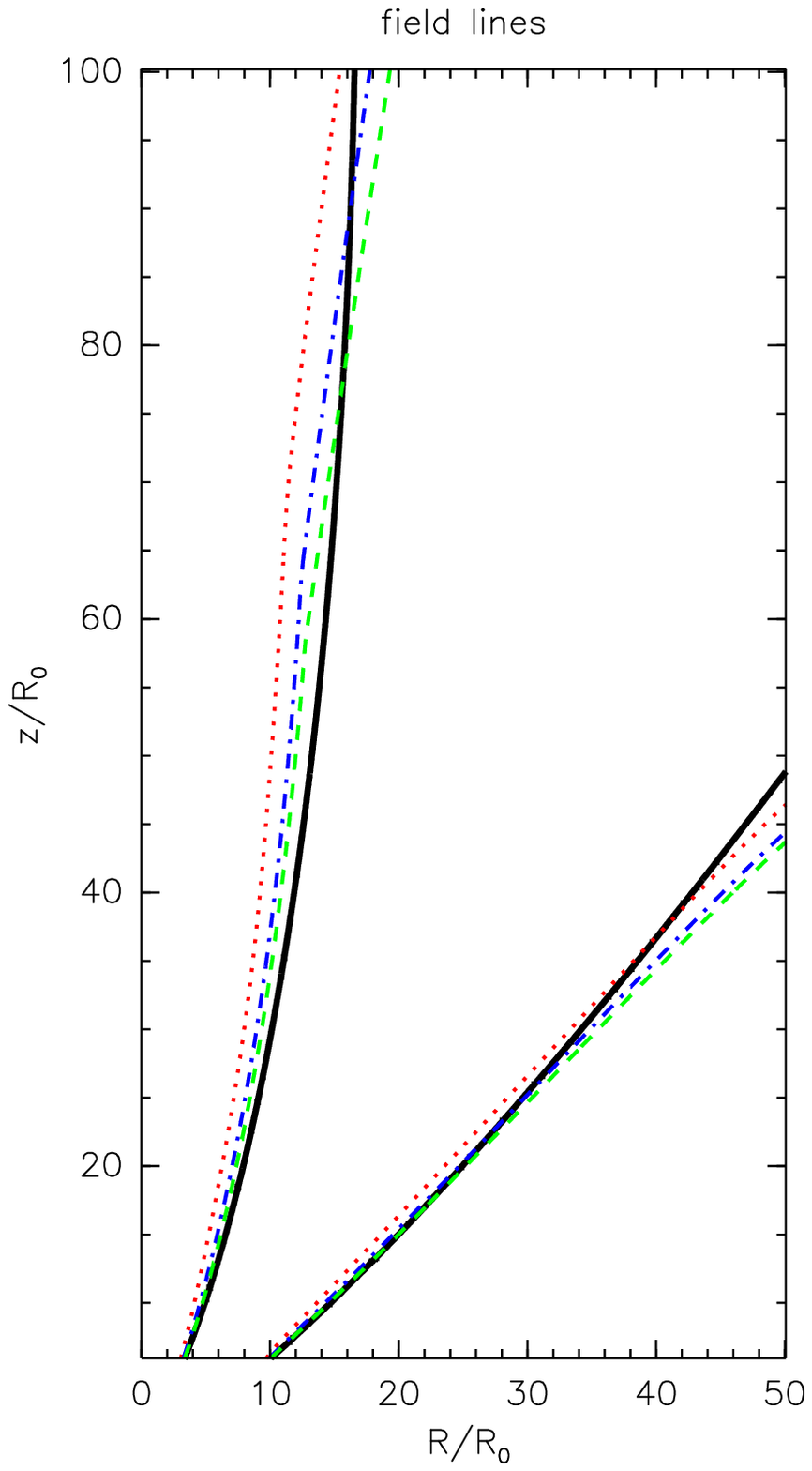}
\includegraphics[width=5.5cm,height=8cm]{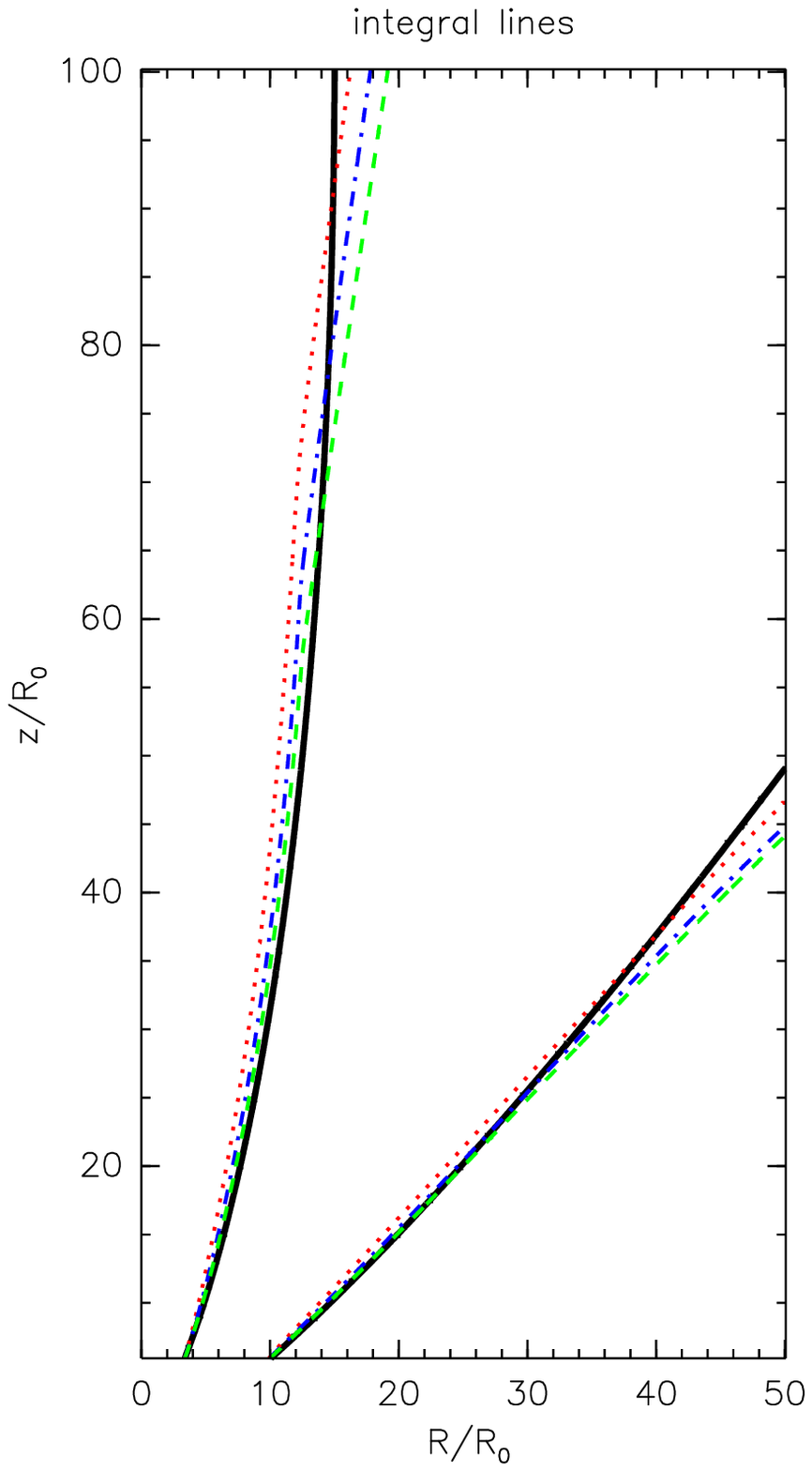}
\caption{Illustration of the effect of {\em numerical} resistivity, i.e. grid
resolution.
{\em Left:} The slow-magnetosonic, Alfv\'enic, and fast-magnetosonic
critical surfaces (from bottom to top).
Different line types represent the final states of simulations with resolution
$128\times 256$ ({\em dotted/red}),
$256\times 512$ ({\em dot-dashed/blue}), and
$512\times 1024$ ({\em dashed/green}).
The {\em solid/black} lines show the initial-state critical surfaces
in the high resolution reference simulation.
{\em Middle:} The
shapes of two different magnetic flux surfaces, i.e. poloidal magnetic field
lines, for the same resolutions and line types as described above.
The solid/black lines represent again the initial state of the
high resolution reference simulation.
{\em Right:} Same as in the {\em Middle} panel, but for the energy $(E)$
integral lines.
}
\label{num_lines}
\end{figure*}
\begin{figure*}
    \includegraphics[height=8cm]{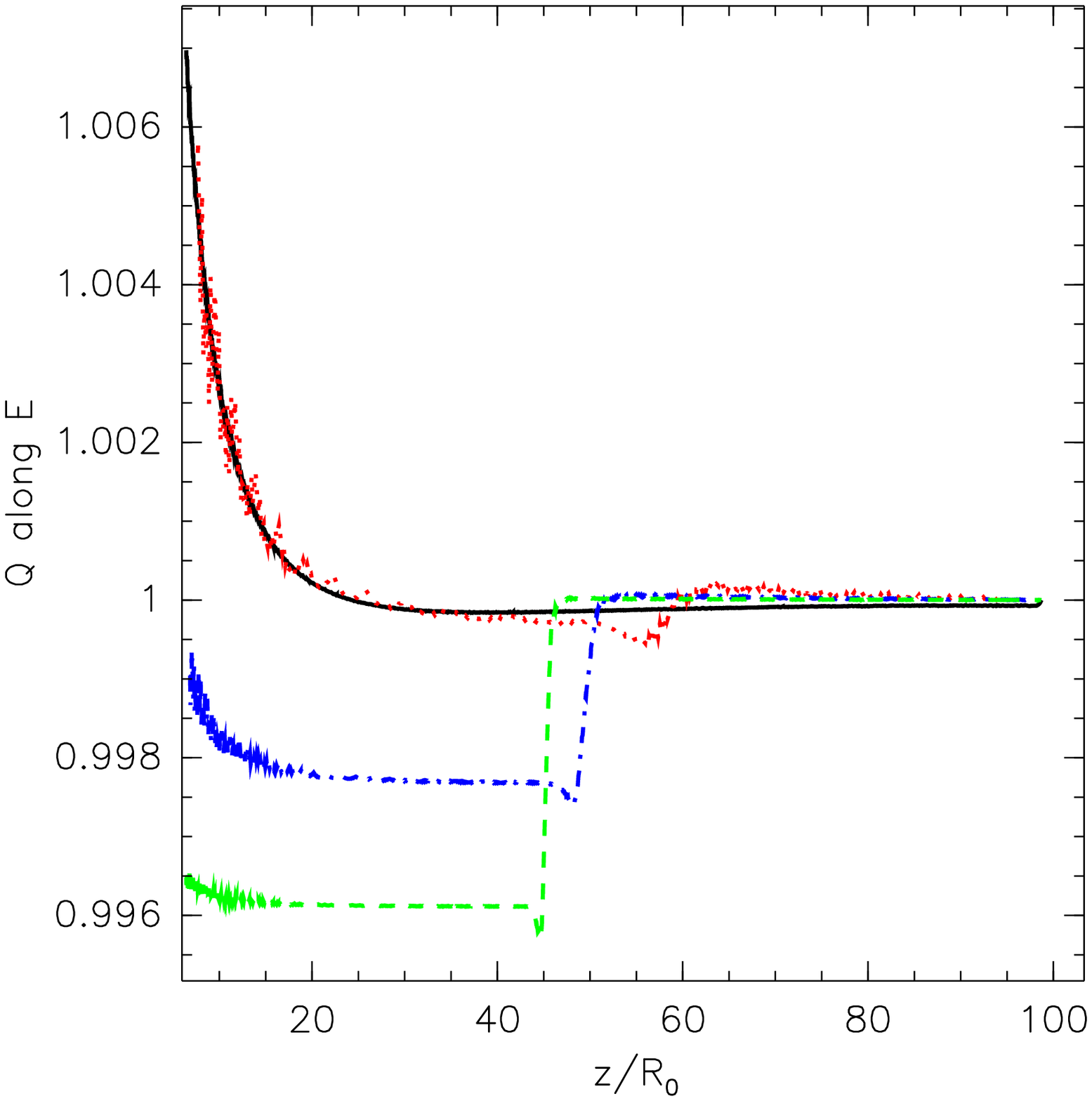}
    \includegraphics[height=8cm]{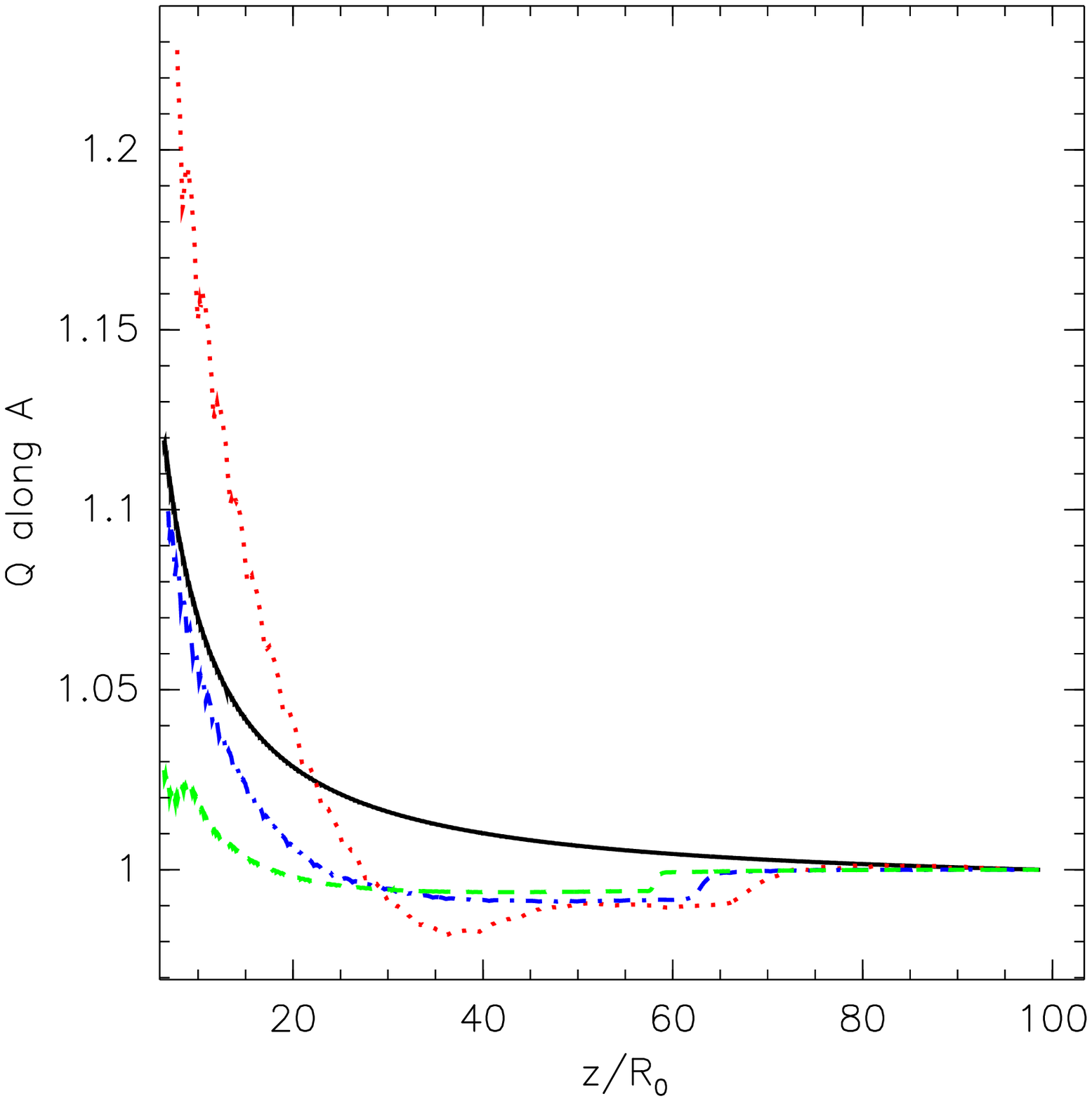}
  \caption{ Illustration of the effect
    of {\em numerical} resistivity, i.e. grid resolution, on the
    alignment of the MHD integrals with the magnetic flux surfaces.
    {\em Left:}
    As an example for the MHD integrals, the entropy $Q$, normalised to
    its value at large distances, is plotted along the inner total
    energy integral line shown in Fig.~\ref{num_lines}, for the initial
    state ({\em solid/black}) of the high resolution reference solution,
    and the final states at resolutions
    $128\times 256$ ({\em dotted/red}),
    $256\times 512$ ({\em dot-dashed/blue}), and
    $512\times 1024$ ({\em dashed/green}).
    {\em Right:} Same as in the {\em Left} panel but here $Q$ is plotted along
    a magnetic field line (instead of the energy integral line).}
\label{num_integrals}
\end{figure*}

\begin{figure*}
    \includegraphics[height=8cm]{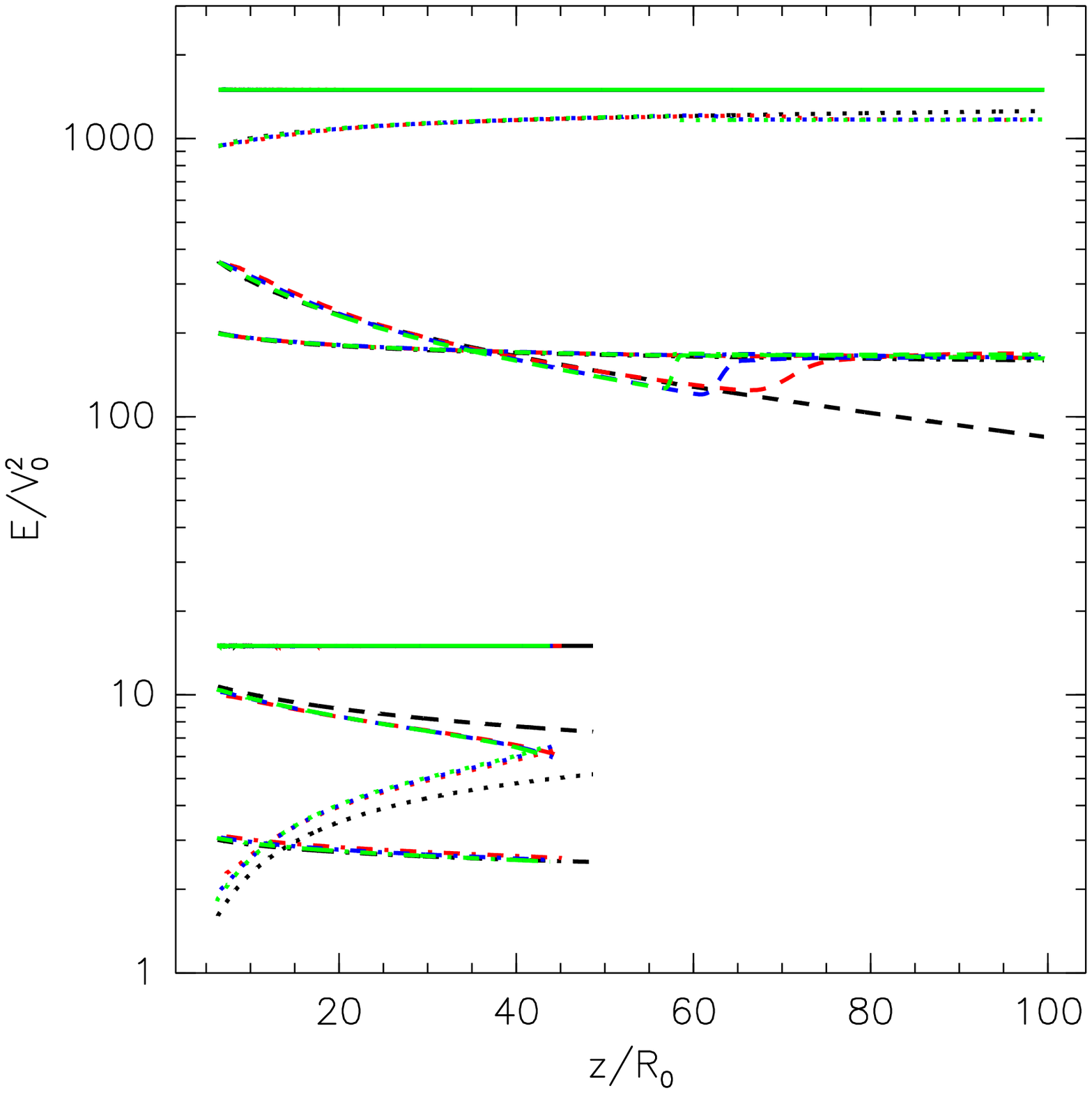}
    \includegraphics[height=8cm]{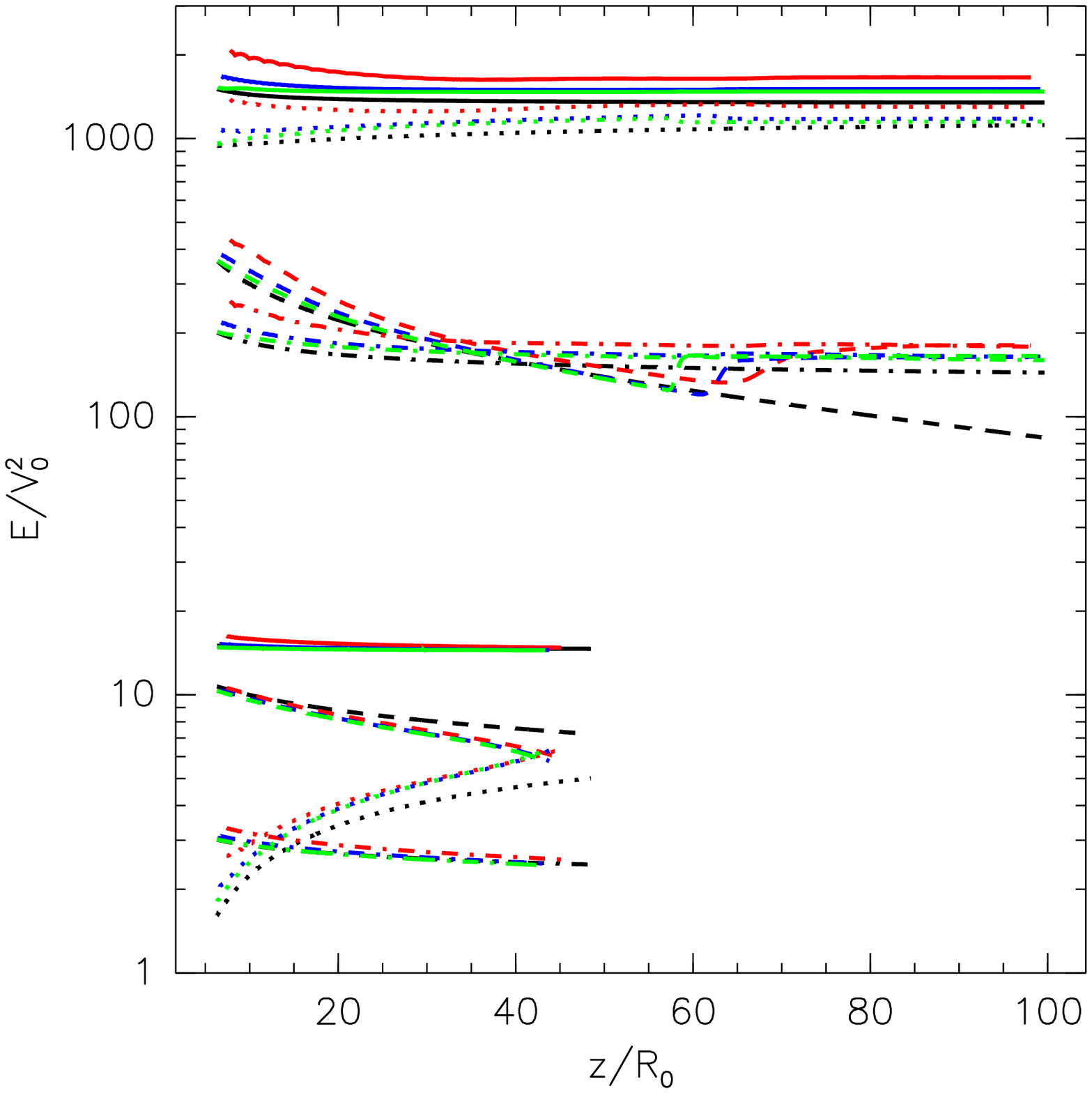}
  \caption{Illustration of the effect of {\em numerical} resistivity,
i.e. grid resolution, on the contributions to the energy
integral $E$. {\em Left:} Split-down of the energy contributions
along the two integral lines shown in Fig.~\ref{num_lines}. The upper set of
curves corresponds to the inner integral line, the lower set of curves to
the outer integral line. Colours code the resolution as in Fig.~\ref{num_lines}.
The different line types represent energy $E$ ({\em solid}),
kinetic energy ({\em dotted}), enthalpy ({\em dot-dashed})
and Poynting ({\em dashed}), respectively. The
gravitational energy is not shown (it is orders of magnitude smaller).
{\em Right} Same as {\em Left} panel, but plotted along the field line instead
of the integral lines.
}
\label{num_eterms}
\end{figure*}
The poloidal velocity field should, in result, also be modified. Initially
$\vec{V}_p\parallel \vec{B}_p$, as demanded for steady ideal-MHD flow.
Therefore, we compute the
new poloidal velocity field, maintaining the magnitude of velocity,
but correcting the direction so that $\vec{V}_p\parallel \vec{B}_p$ holds.

The modified initial magnetic field is presented in Fig.~\ref{pocet},
together with the modified density isocontours and positions of the critical
surfaces.

For boundary conditions we use symmetry conditions on the
rotation axis, and outflow conditions on the outer $R$ and $Z$ boundaries.
On the lower boundary $Z=Z_{\rm min}$ 
(we take $Z_{\rm min}=6 R_0$ in all simulations)
we fix the values for six physical quantities,
namely density, three velocity components, the azimuthal and one poloidal magnetic
field component (the other one is given from $\nabla\cdot B=0$). The pressure/internal
energy boundary condition is kept fixed only in the super-sonic region (where the
poloidal velocity is larger than the speed of sound), otherwise it is
extrapolated from the flow onto the boundary.

More specifically, in our simulations the $Z$-component of the magnetic field
in the first ghost cell is set from the analytical solution, and
the radial component is obtained from the divergence-free condition. Therefore, the
magnetic flux along the boundary is fixed at all times. In the second ghost cell the
$Z$-component of the magnetic field is extrapolated linearly, to allow for
change in the field line shape, when the radial component is again obtained from
$\nabla\cdot B=0$.

In our computations we used various resolutions and sizes of the
computational domain. Here we present
the results for the resolution
$R\times Z=(256\times512)$ grid cells $=([0,50]\times [6,100])R_0$, in the uniform
grid. Results comply with the solutions for one fourth, one
half and double of this resolution, which we also computed and presented,
when needed for direct comparison.

\subsection{Ideal-MHD integrals}
It can be shown, that
steady, axisymmetric, ideal-MHD polytropic flows
conserve five physical quantities
along the poloidal magnetic field lines (Tsinganos 1982).
These so called {\em integrals} are
the mass-to-magnetic-flux ratio $\Psi_A$,
the field angular velocity $\Omega$,
the total angular momentum-to-mass flux ratio $L$,
the entropy $Q$, and
the total energy-to-mass flux ratio $E$ (henceforth we call
the latter integral energy for brevity).
These integrals are given as
\beq
  \Psi_A = \frac{4 \pi \rho V_p}{B_p} \,,
\eeq
\beq
  \Omega = \frac{V_\phi}{R} - \frac{B_\phi}{B_p} \frac{V_p}{R}\,,
\eeq
\beq
  L = R V_\phi - \frac{R B_\phi B_p}{\mu_0 \rho V_p} \,,
\eeq
\beq
  Q = p/\rho^\gamma \,,
\eeq
\beq \label{energy_int}
  E = \frac{V^2}{2} + \frac{\gamma}{\gamma-1} \frac{P}{\rho}
  + \frac{ B_\phi \left( B_\phi V_p - B_p V_\phi \right)}{\mu_0 \rho V_p}
  -\frac{\cal GM}{r}
\,.
\eeq
The various contributions of the energy $E$ correspond to the various
terms on the right hand-side of Eq.~(\ref{energy_int}). From left to right,
the kinetic, enthalpy, Poynting, and gravity terms
can be recognised.

The degree of alignment of the lines on the poloidal plane where the above
quantities are constant together with the
poloidal magnetic field lines
can be used as a test on how close to a
steady-state is the final result of a simulation.

\section{Ideal-MHD simulations and numerical resistivity}
Ideal-MHD ($\eta=0$) numerical simulations with the same setup and using
the same code have been performed in GVT06.
The density isocontour plots for ideal MHD simulations in different times
are presented in Fig.~\ref{idealdens}, for illustration of the relaxation process.

In the early stage of simulation, the initial conditions, especially the setup near
the symmetry axis, affect the outflow time-evolution. Up to few thousands of
Courant time steps, the solutions might look moderately different, before
the relaxation towards the stationary state. Finally, after few ten thousand
Courant time-steps (or in higher resolutions after
few hundred thousands) the simulations give similar result as the initial
state. This solution is stationary (not quasi-stationary)
as we did not note virtually any change in a time ten times longer than the time
needed to reach the stationary state. This time is equal to five millions of 
Courant times steps, or $2500 R_0/V_0$, 
when expressed in normalised units.

The initial setup near the axis, which may introduce big differences
in the initial state (compared to the self-similar solution),
does not affect the reached final state so much as
the corresponding part of the boundary $Z=Z_{\rm min}$ near the disk surface.
These ideal MHD results have been extensively discussed in GVT06.

For our purpose here, to use them as a reference solution for the resistive runs,
the problem of numerical resistivity should be addressed.

In ideal-MHD numerical simulations, a current sheet, which forms at any
discontinuity of the magnetic field, should not diffuse away. Also, magnetic field
lines should not reconnect through such a sheet. However, since in a
finite-difference scheme computations can not resolve features
smaller than grid cells, {\em numerical} reconnection occurs, cf. Hawley \& Stone (1995).

To check the effects of the numerical resistivity, we compared our ideal-MHD
simulations performed in various resolutions. These were
$R\times Z=(128\times256)\,, (256\times512)$
and $(512\times1024)$ grid cells, in identical setups.
In Fig.~\ref{num_lines} the effect of grid resolution
(i.e. the effect of the numerical resistivity)
on the position of the critical MHD surfaces,
on the shape of the field lines, and on integral lines, is shown.
For increasing numerical resistivity (i.e. for lower resolution), the critical
surfaces move downstream. However, towards large cylindrical radii $R$ and small
height $Z$, the effect of the boundary conditions becomes important
and the critical surfaces bend towards the disk, as noted in GVT06.
Also, for increasing numerical resistivity
the field lines tend to straighten out and move to smaller cylindrical radius.
However these differences are insignificant.

In Fig.~\ref{num_integrals} the entropy integral $Q=p/\rho^\gamma$ is shown,
along the same field lines as in Fig.~\ref{num_lines},
normalised to its value at large distance.

In the ideal MHD case, all integral lines should coincide.
Fig.~\ref{num_integrals} illustrates that
the initial state does not show perfect alignment of the integrals.
This is expected, since the initial state is a modified exact
solution of the ideal-MHD equations.
For the final state the integral lines at
any resolution are better aligned than in the initial state.
As expected, the alignment of the integrals is
better for lower numerical resistivity, i.e. higher resolution.
In contrast to the analytical solution, the
numerical solution features a
shock (see a related discussion in \citealp{M08}).
As expected, the entropy jumps across the shock, as illustrated
in Fig.~\ref{num_integrals},
but is otherwise constant along the energy integral line.

Further, for ideal-MHD steady flows the integrals fall exactly on magnetic flux
surfaces, i.e. field lines. However, as shown in the
right panel of Fig.~\ref{num_integrals}, the presence of
numerical resistivity changes this situation. While the integrals stay
well-aligned among themselves, the alignment with the magnetic
field lines is poor for low resolution ($128\times 256$) and becomes
almost perfect for high resolution ($512\times 1024$) runs.
In fact, while the integrals re-align while
the simulation progresses, the field lines diffuse away from the
corresponding integral lines over time. However, for all numerical
resolutions in our simulations this process eventually comes to halt 
and a stationary state is reached.

Not only are the integrals aligned very well, but also the
individual contributions to the energy $E$ along the integral
lines are very similar across different numerical diffusivities as
illustrated in Fig.~\ref{num_eterms}. Apart from resolution effects at
the shock along the inner integral lines, all models follow similar
trends and converge to the same profiles.

Again, plotting the same quantities along field lines, as shown in the right
panel of Fig.~\ref{num_eterms} shows large
spread across different numerical resistivities. Higher values of
resistivity show larger changes along the field line, as well as do
field lines further in.

We conclude that numerical resistivity does effect the solution,
but in a smooth manner. For reasonable resolutions (grid cells small
enough compared to the characteristic length of the problem), it does not
challenge the solution in our setup.

\section{Resistive-MHD simulations}
To investigate the resistive-MHD behaviour of such outflows,
we set the resistive-MHD numerical simulation for NIRVANA with the
same initial and boundary conditions as for the ideal-MHD simulations
presented in the previous section.
\begin{figure*}
    \includegraphics[width=5.5cm,height=8cm]{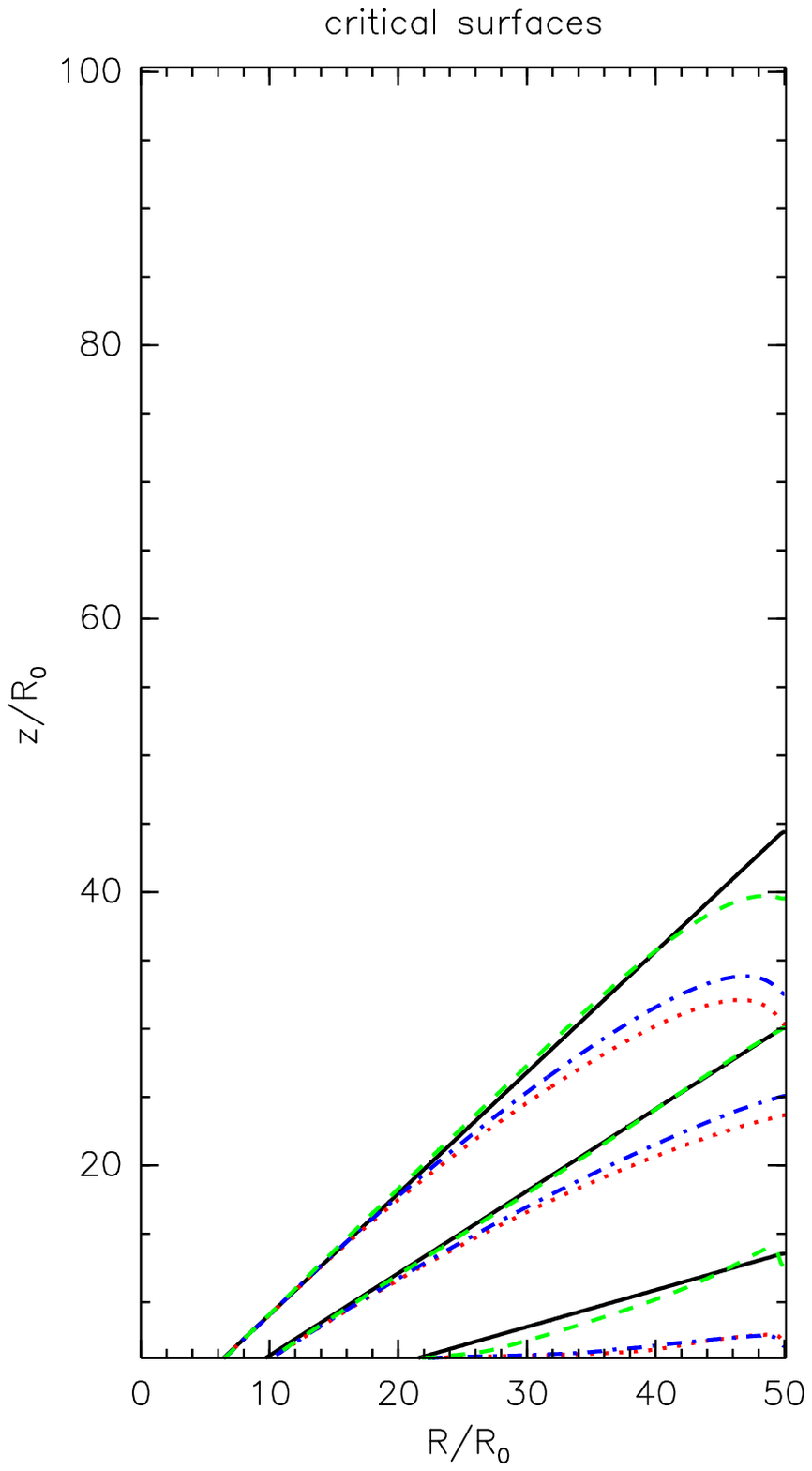}
    \includegraphics[width=5.5cm,height=8cm]{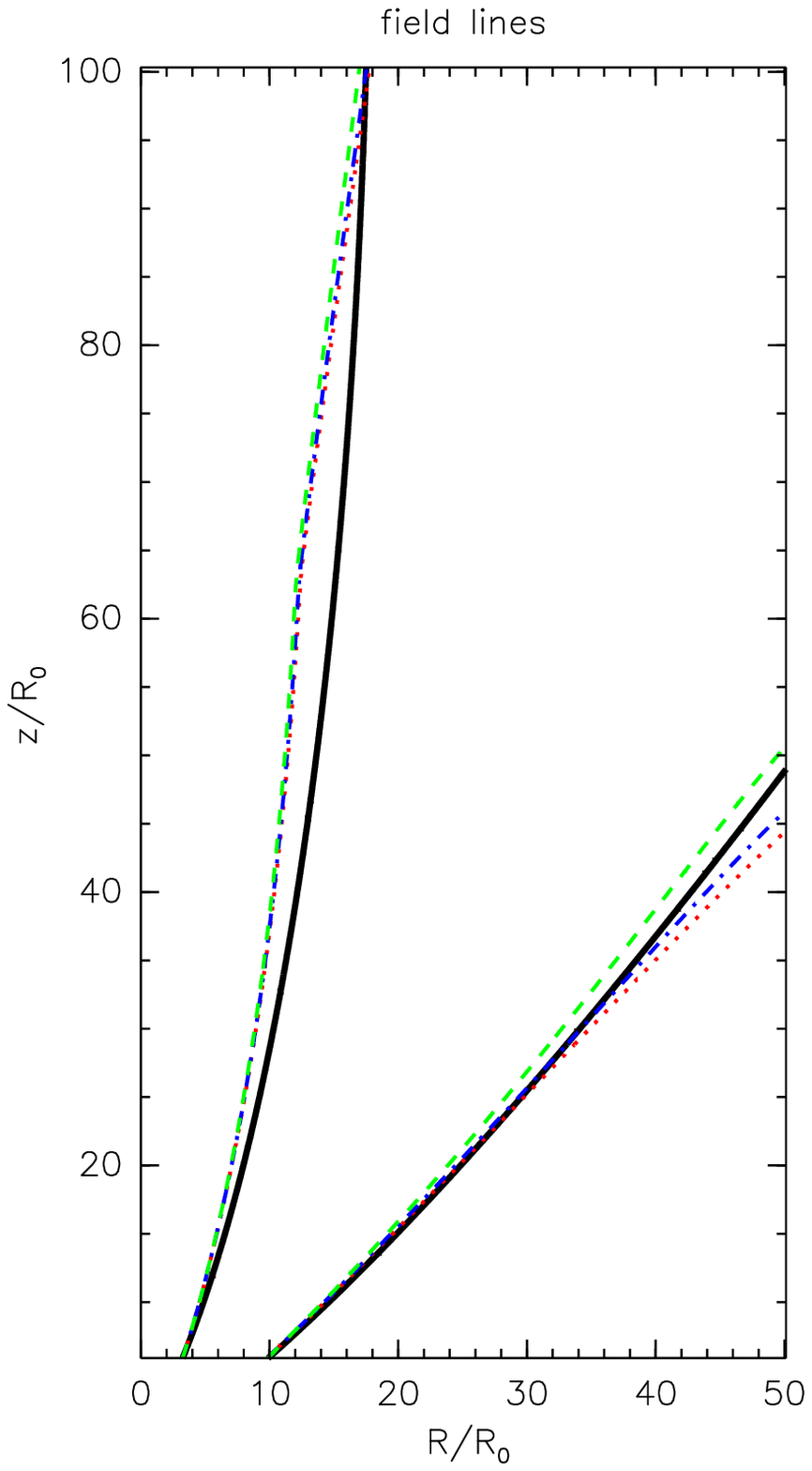}
    \includegraphics[width=5.5cm,height=8cm]{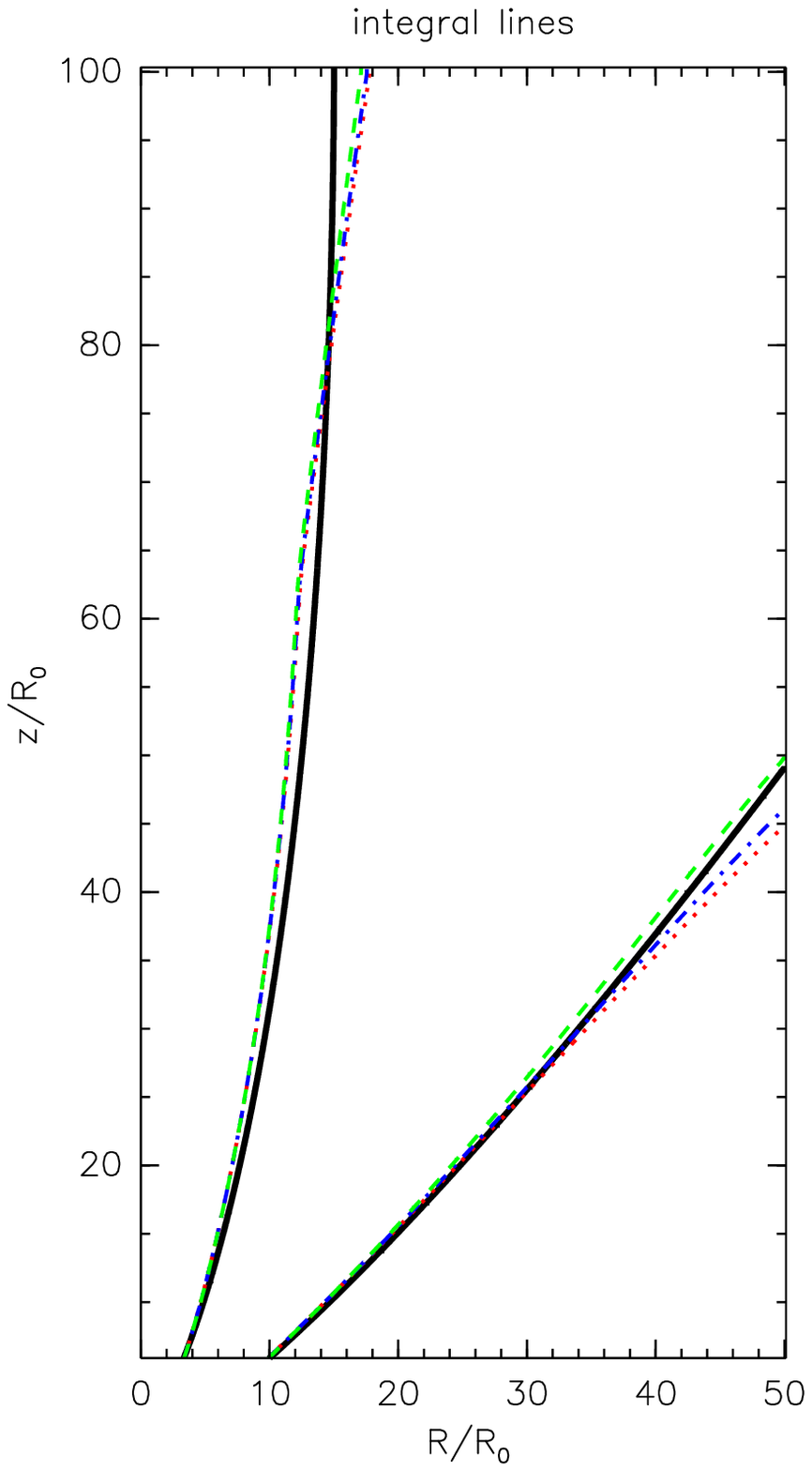}
  \caption{Illustration of the effect of {\em physical} resistivity.
{\em Left:} The slow-magnetosonic, Alfv\'enic, and fast-magnetosonic
critical surfaces (from bottom to top).
Different line types represent the final states of simulations at resolution
$R\times Z=(256\times512)$ grid cells,
with the physical magnetic resistivity $\etahat=0$ ({\em dotted/red}),
$\etahat=0.03$ ({\em dot-dashed/blue}) and $\etahat=0.15$ ({\em dashed/green}).
The {\em solid/black} lines show the initial state critical surfaces.
{\em Middle:} The
shapes of two different poloidal magnetic field lines, for
the same resolutions and line types as described above.
The {\em solid/black} lines represent again the initial state.
{\em Right:} Same as in the {\em Middle} panel, but for the energy
$(E)$ integral lines.
}
\label{res_lines}
\end{figure*}
\begin{figure*}
    \includegraphics[height=8cm]{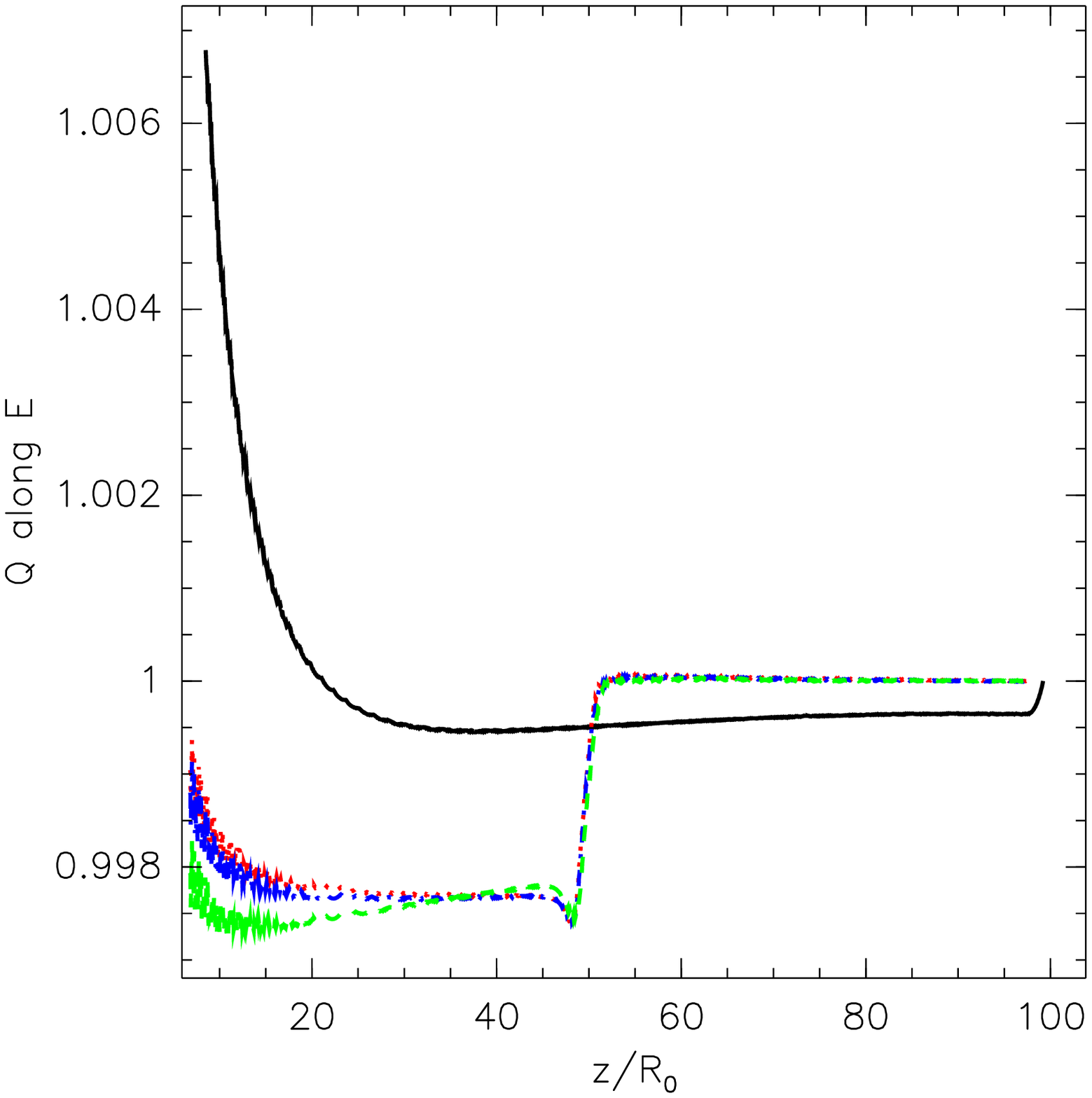}
    \includegraphics[height=8cm]{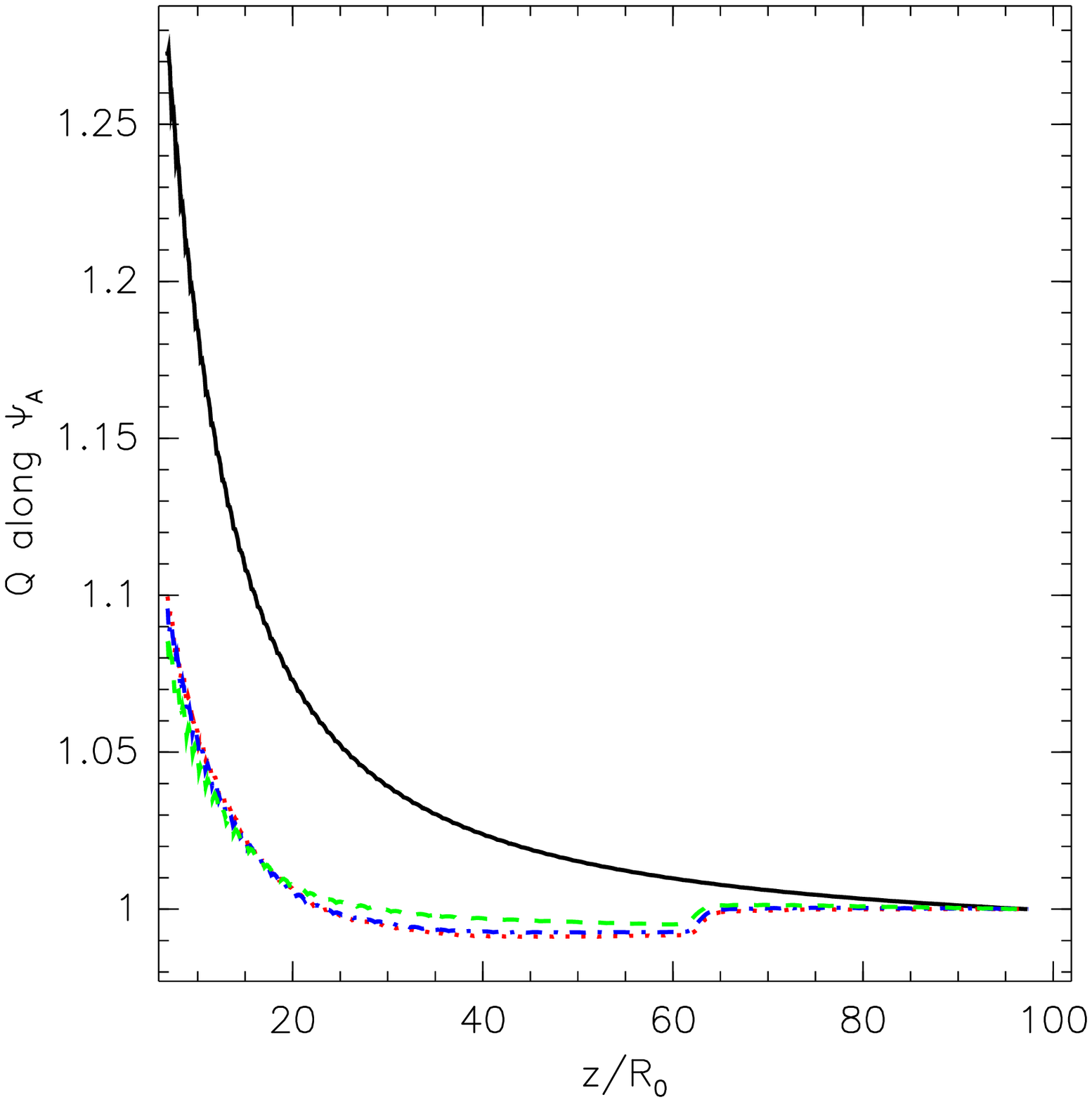}
  \caption{Illustration of the effect of {\em physical} resistivity on
    the alignment of MHD integrals and magnetic flux surfaces.
    {\em Left:} As an example for the MHD integrals the entropy $Q$,
    normalised to its value at large distances, is plotted along the
    inner energy integral line shown in Fig.~\ref{res_lines} for
    the initial state ({\em solid/black}), and final states for
    $\etahat=0$ ({\em dotted/red}), $\etahat=0.03$ ({\em dot-dashed/blue})
    and $\etahat=0.15$ ({\em dashed/green}) for runs with resolution
$R\times Z=(256\times512)$ grid cells.
{\em Right:} Same as in the {\em Left} panel but plotted
along the innermost field line (instead of the energy integral line).
}
\label{res_integrals}
\end{figure*}
\begin{figure*}
    \includegraphics[height=8cm]{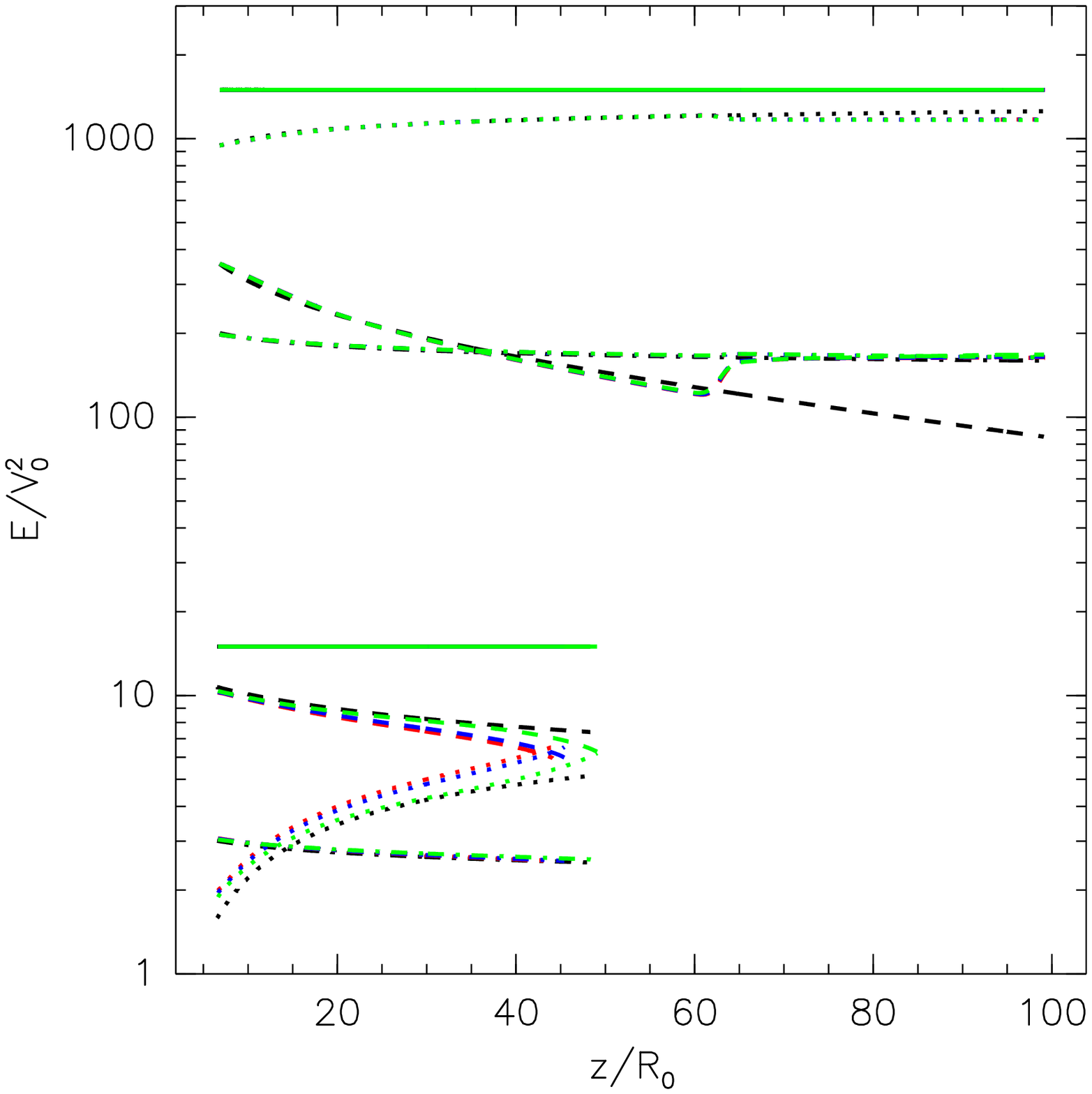}
    \includegraphics[height=8cm]{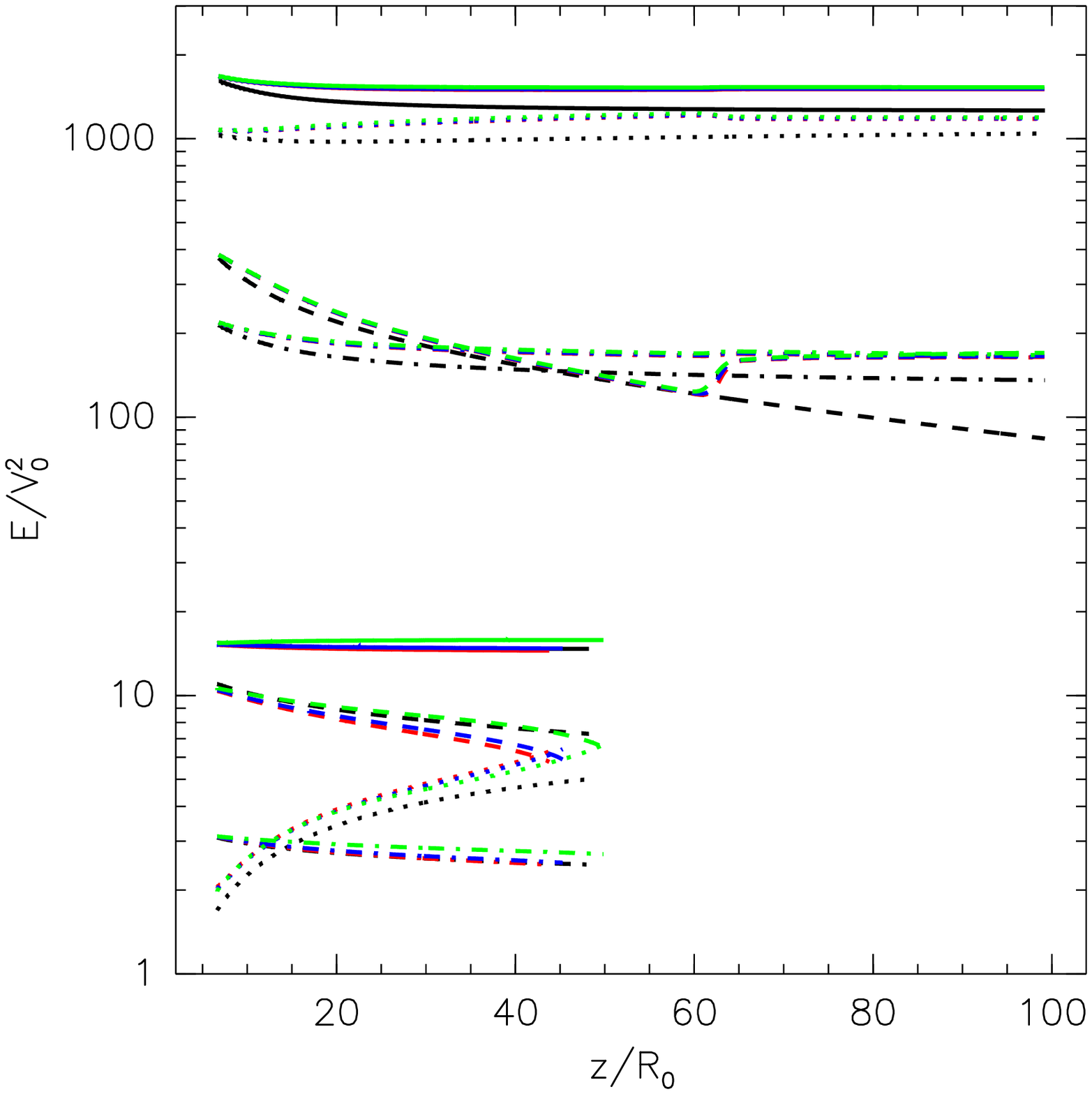}
  \caption{
    Illustration of the effect of {\em physical} resistivity on
   the contributions to the energy integral $E$.
   {\em Left:}
   Split-down of the energy contributions along the two integral
   lines shown in Fig.~\ref{res_lines}. The upper set of curves
   corresponds to the inner integral line, the lower set of curves to
   the outer integral line. Colours code the physical magnetic diffusivity
   as in Fig.~\ref{res_lines}. The different line types represent energy
   $E$ ({\em solid}), kinetic energy ({\em dotted}), enthalpy ({\em
   dot-dashed}) and Poynting flux energy ({\em dashed}).
   The gravitational energy is not shown (it is orders of magnitude
   smaller).
   {\em Right:} Same as {\em Left} panel, but plotted
   along the field line instead of the integral lines.}
\label{res_eterms}
\end{figure*}
The magnetic diffusivity is set to constant throughout the computational
box. It would
be possible to model the diffusivity as proportional to a product of a
characteristic length (e.g. the cylindrical distance) with a
characteristic velocity of the problem
(e.g. the Alfv\'en velocity or the sound speed). Alternatively, we could model an ambipolar diffusivity
as a non-constant scalar $\eta = V_A^2 \tau_{ni}$, where $V_A$ is the Alfv\'en speed 
and $\tau_{ni}$ is the neutral-ion momentum exchange time.
In the special case that the magnetic field and current density are orthogonal
(as is the case close to the disk where the poloidal field dominates)
this is an exact expression (e.g., Balbus \& Terquem, 2001).
However, in this study we opted for the simplest case of a constant $\eta$.
It is expected that, at least for relatively low values of $\eta$, the exact
prescription of diffusivity should not significantly affect the flow.
As noted in FC02, where diffusivities $\eta \propto  \rho^{1/3}$
were examined,\footnote{
These cases correspond to
$\eta \propto C_s \propto \rho^{(\gamma-1)/2} \propto \rho^{1/3}$
for $\gamma =5/3$ and $P/\rho^\gamma$ being a global constant
(and not constant along the flow as in the self-similar model of V00).}
differences introduced by relating the resistivity to density are small.
Variable resistivity cases in which the exact prescription may
significantly affect the results in the high resistivity limit,
will be examined in another connection.

The nature of magnetic diffusivity which we introduced in our simulations depends
on the specific circumstances valid for the investigated flow. We treat it as an
"effective magnetic diffusivity", without discussing its physical origin, an issue that would
be beyond the scope of this paper. The most obvious
case could be magnetic turbulence, which would extend from the disk to the disk
corona immediately above the disk. Since we do not treat the disk, but take it as a
boundary condition here, we can not treat the resistivity self-consistently.
Therefore we take it as a free parameter. 

We performed a study of magnetic resistivity in our simulations. At
first, preparatory work has been done, with level
of numerical magnetic diffusivity tracked by decreasing the parameter $\etahat$ until
there was no effect on the solutions (i.e. until they became identical to the
ideal MHD solutions, obtained by the code). In our setup here it showed to be of the order
of $\etahat\sim 0.001$.

Then we started increasing the magnetic diffusivity. The solution remained
similar in character to the ideal-MHD one until some
threshold critical magnetic diffusivity $\etahat_c=0.15$ has been reached.
For $\etahat >\etahat_c$ the solution
changed abruptly, not resembling the initial condition anymore. Similar behaviour
has been reported in FC02 who also refer to a critical magnetic diffusivity,
but in their study a comparison to some analytical solution was not possible.\footnote{
In FC02 the setup was motivated by reasons of comparison with \cite{OP97a}
ideal MHD {\em numerical simulations}, which reach well defined quasi-stationary
state, needed for successful comparisons, but are not necessarily stationary even
in the ideal MHD regime. Quasi-stationarity of the solutions in FC02 has been
defined rather by a "rule of thumb", when here the solution reaches well defined
stationary state, which is possible to relate and compare to the analytical solution.}
More importantly, FC02 ignored the resistive term in the energy Eq.~(\ref{enn}),
and thus they could not observe a modification in the flow caused by the energy dissipation
as we do here. Their critical resistivity has to do with the diffusion of the magnetic field;
as a result we cannot directly compare the two works.

Analysis and direct comparison of the data for the resistive runs with the ideal-MHD
analytical solutions can be performed only when the solutions do not depart
largely from the stationary ideal-MHD ones.
In such case the integrals along the similar lines can be compared.
For the large resistivity, as the solutions differ significantly,
a separate study of validity and stationarity of the new solution is required.

Therefore, here we concentrate on the solutions not departing significantly from the
ideal-MHD solutions.

The positions of critical surfaces and field line shapes for the different physical
resistivities (but for single resolution) are shown in Fig.~\ref{res_lines}.
As was the case with the numerical resistivity,
the critical surfaces move downward the flow with increasing resistivity.
This effect is more prominent here, meaning that $\etahat=0.03$
gives an upper limit for the numerical resistivity
of the $R\times Z=(256\times512)$ resolution.

Also, the field lines tend to straighten out.
However, field lines close to the axis seem to be little, if at all, affected by
the resistivity, leading to the conclusion that the flow is not modified along them.
This is different than was the case for the numerical resistivity.

The MHD integrals (see Fig.~\ref{res_integrals}) are not well conserved along the
flux surfaces, as was observed also for the numerical resistivity.
However, the misalignment is not enlarged.

The evolution of individual contributions of the energy (see Fig.~\ref{res_eterms})
shows a clear trend along the flux surfaces.
For increasing resistivity, the Poynting energy and enthalpy increase, while the
kinetic energy decreases. The differences are not big, though,
especially in the outer field lines. Again, the field lines close to the axis are
special in the sense that
all the curves fall atop of each other, indicating that the energy there is
independent of the physical resistivity.

The curves for $\etahat=0$ and $\etahat=0.03$ are almost identical, what amounts to the
conclusion of our preparatory work mentioned above, about the order of numerical
magnetic diffusivity. It is confirmed now that $\etahat=0.03$ is a low physical
magnetic diffusivity case.

As in the numerical diffusivity cases,
not only are the integrals aligned very well, but also the
individual contributions to the energy $E$ along the integral
lines are very similar across different diffusivities, as
illustrated in Fig.~\ref{res_eterms}.

In general, the time evolution of the NIRVANA solutions for a low magnetic diffusivity
parameter $\etahat\leq 0.15$ does not differ much from the ideal MHD evolution.
It only takes more computational timesteps to reach the stationary state, as
the diffusive timestep now adds to the total timestep.

After a few hundred thousands (or few millions, for larger resolutions) Courant
timesteps of the relaxation process, the outflow
reaches the stationary state, which is similar to the initial state.
The relaxation is not dramatic, as the flow is without strong shocks, and connection
of the initial
condition to the boundary conditions is smooth. Such evolution is expected for given
initial conditions.

Here we also presented solutions for the magnetic diffusivity close to the highest
one which does not change the solutions dramatically, $\etahat=0.15$. For larger
magnetic diffusivity than this threshold, the solutions seem to depart
significantly from the initial condition. Therefore, it is not possible to describe
these solutions simply comparing the integrals along the same lines, as we did in the
analysis above, because the geometry of the solutions is completely different.

\begin{figure}
\hspace{1.28cm}\includegraphics[width=6.01cm, height=1.cm]{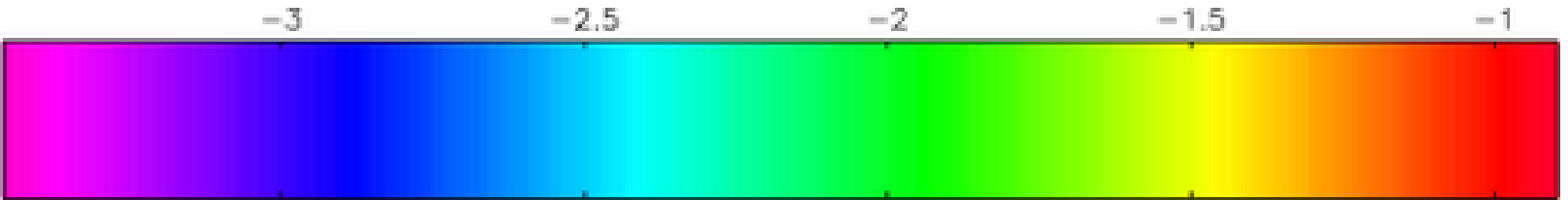}
\includegraphics[width=7.5cm, height=12cm]{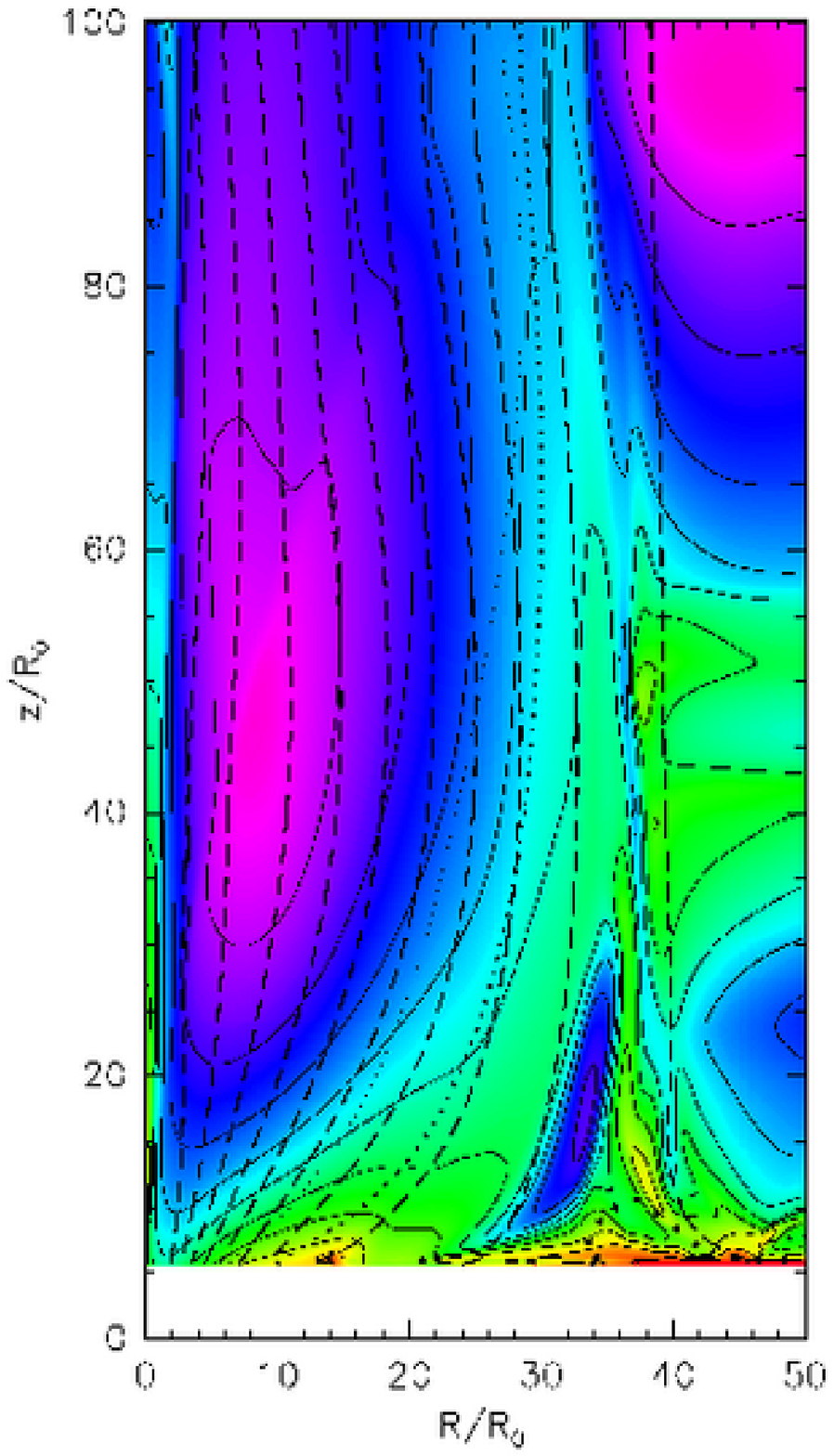}
\caption{ Illustration of the effect of super-critical diffusivity on the
flow structure. This simulation with $\etahat=1.5$ never reaches a
steady state, but remains highly time-variable. The solid
lines represent logarithmically spaced isocontours of density. It is also shown 
in colour grading. The dashed lines show poloidal fieldlines.
The dotted lines indicate, from top to bottom, the position of the fast-magnetosonic
({\em small dots}), the Alfv\'en, and the slow-magnetosonic surface ({\em
large dots}), respectively.
}
\label{eta1a}
\end{figure}

Preliminary results for the high resistivity regime in our setup
are shown in Fig.~\ref{eta1a}.
The flow with $\etahat=1.5$ is examined, i.e.
with diffusivity 50 times larger than the typical "low
magnetic diffusivity" value. The solution departs significantly from the
ideal MHD case, and seems to show some periodicity in time-evolution. This 
result of the radical deviation of the solution from the stable ideal 
MHD jet solution may have some interesting astrophysical implications. For example, 
one possibility is that if somehow the resistivity drastically increases in the 
disk (e.g., due to some instability), a well formed and behaved jet ceases to exist, giving its place to a more 
erratic outflow. However, this investigation is beyond the scope
of the present paper, where we check the behaviour and stability of the ideal MHD
solutions with the inclusion of the resistivity for the particular problem setup. 
Instead, it will be examined in a following paper.
\begin{figure}
\hspace{-5cm}\includegraphics[width=18.0cm]{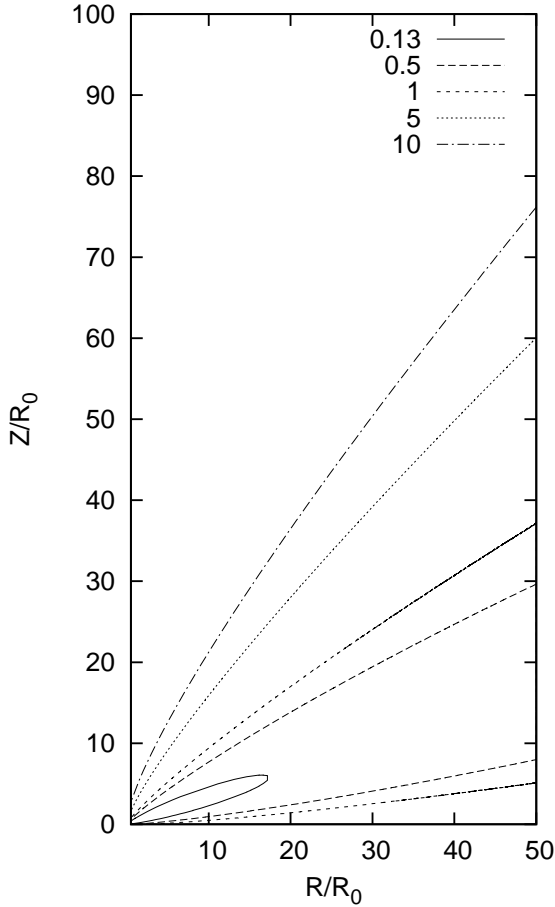}
\caption{Value of $0.5\beta (VR/ V_0 R_0)$ for the analytical solution of V00.
This quantity gives the critical value $\etahat_c$ that corresponds to $\rbeta =1$.
}
\label{analyt_cont}
\end{figure}

\section{A critical value of the diffusivity in MHD jets}

The resistive effects in a magnetohydrodynamic flow have been traditionally
quantified by using the magnetic Reynolds number $R_{\rm m}$,
the dimensionless ratio of the advection and diffusion parts
in the induction Eq.~(\ref{faraday}).
In a case of a jet the advection velocity is the flow speed $V$, while a
characteristic scale measuring the distance at which the various
quantities vary substantially is the cylindrical radius $R$. Thus, the
Reynolds number is, 
\beq
R_{\rm m}=\frac{VR}{\eta}\,,
\eeq
and magnetic diffusion is unimportant for $R_{\rm m}\gg 1$.

However, the resistivity affects also the energy transport, Eq.~(\ref{enn}), through the
Joule heating term.
The ratio of non-resistive terms in this equation over the
Joule heating term gives another important dimensionless number, 
$\rbeta \equiv (PV/R)/(\eta B^2 / \mu_0 R^2)$, or, in terms of the
plasma beta $\beta= 2 \mu_0 P / B^2$,
\beq
\rbeta=\frac{\beta}{2} \ \frac{VR}{\eta} = \frac{\beta}{2}  R_{\rm m} \,.
\eeq
Energy dissipation is important for $\rbeta \la 1$.
Clearly, for magnetised jets with $\beta<2$ it is $\rbeta < R_{\rm m} $.
If $R_{\rm m}<1$ then $\rbeta <1 $ as well, meaning that resistivity effects are
important in both, the induction and the energy equation.
However, there is a possibility to have $R_{\rm m}\gg1 $ and $\rbeta \la 1 $.
These inequalities define a regime where resistivity affects the energy,
but not the induction equation.

Taking flows with progressively larger diffusivity, 
as the numerical experiments that we have performed in this study, we will first
reach a point where $\rbeta \approx 1 $, while still the Reynolds number is $R_{\rm m}\gg 1$.
From this point on, the flows will significantly differ
from the ideal-MHD initial conditions, something which is connected to the
critical diffusivity that we observe.
By writing $\eta=\etahat R_0 V_0 $ we find that the condition $\rbeta = 1 $
gives a critical value
\beq
\etahat_c=\frac{\beta}{2} \ \frac{V}{V_0} \ \frac{R}{R_0} \,.
\eeq
This quantity for the analytical solution of V00 is shown in Fig.~\ref{analyt_cont}.
Inside the region $Z/Z_0 > 6$ that we simulate,
the minimum value of $0.5\beta (VR/ V_0 R_0)$ is $0.13$
(near the point $R/R_0 \approx 17$, $Z/R_0=6$). This means that only
for $\etahat > 0.13$ resistive effects start to play a role --
their influence is more important (at least initially)
near the point $R/R_0 \approx 17$, $Z/R_0=6$.
This value is very close to the numerically evaluated critical value $\etahat_c=0.15$.

\begin{figure*}
\includegraphics[width=6.0cm,height=8.cm]{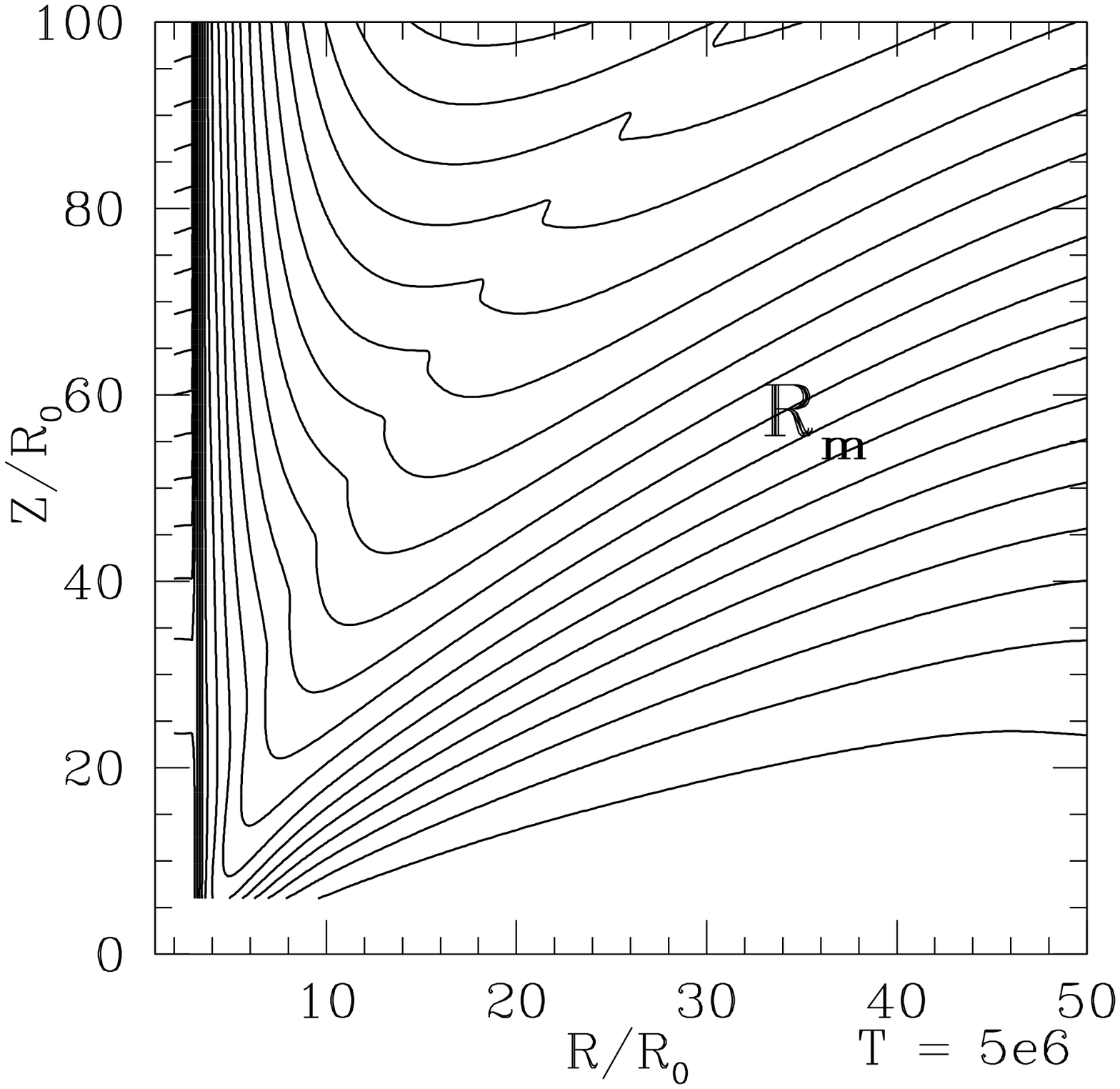}
\includegraphics[width=8.0cm,height=8.cm]{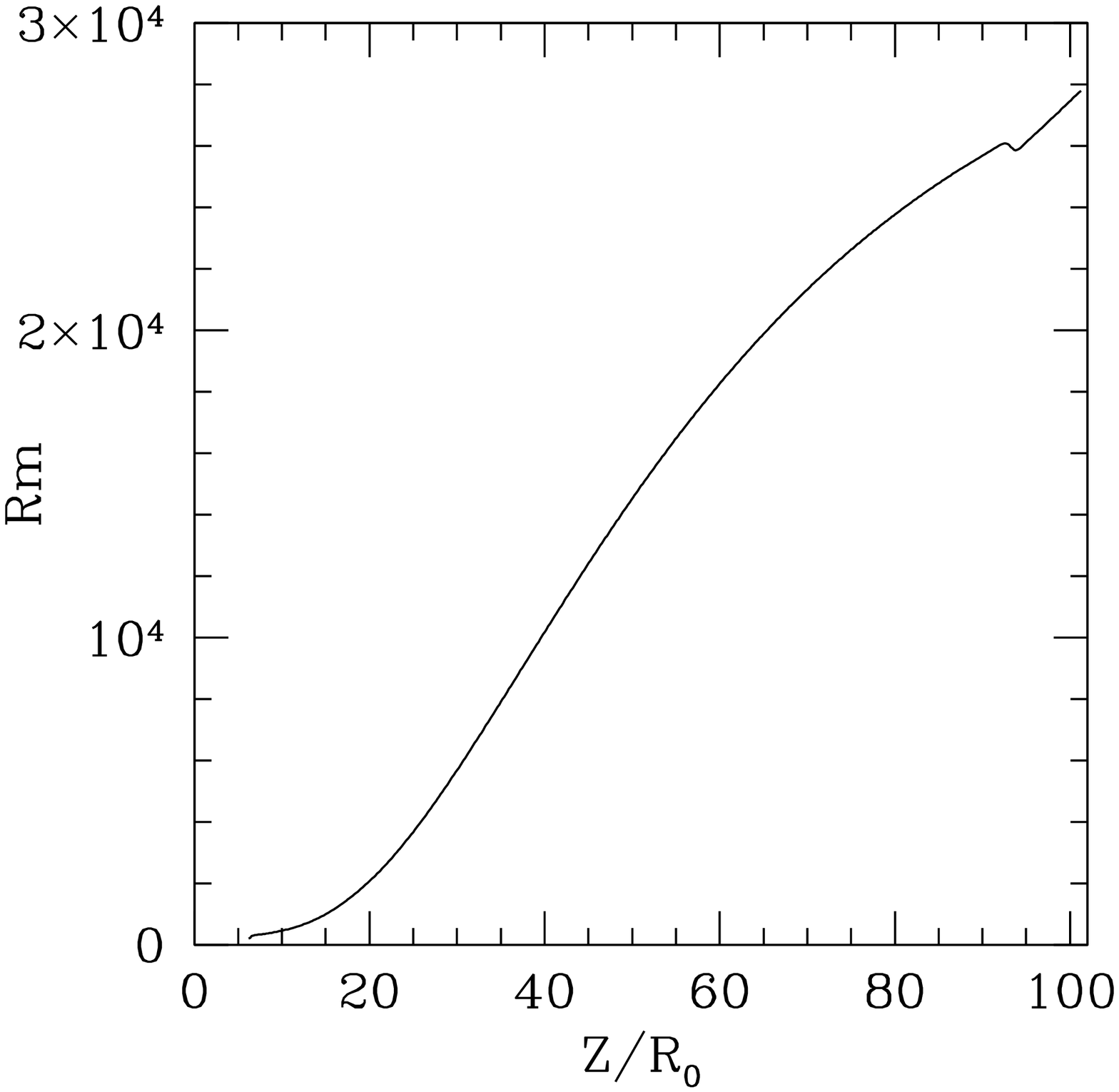}
\caption{The magnetic Reynolds number $R_{\rm m}=VR/\eta$ for $\etahat=0.03$ in
the computational box when $R\times Z=256\times 512$ grid cells, for the physical
domain $R\times Z=([0,50]\times [6,100])R_0$. In the {\em Left} panel
the isocontours of $R_{\rm m}$ are shown. 
The values of $R_{\rm m}$ are from $1500$ to $30000$ in $20$
contour lines, increasing linearly from bottom (near the disk surface) to top. 
In the {\em Right} panel we show $R_{\rm m}$ for the same case,
as a function of height above the disk, at a constant
cylindrical distance corresponding to the middle of the computational box. 
The minimum $R_{\rm m}$ along this line is $335$.
}
\label{rmags2}
\end{figure*}

\begin{figure}
\hspace{1.03cm}\includegraphics[width=4.82cm, height=1.cm]{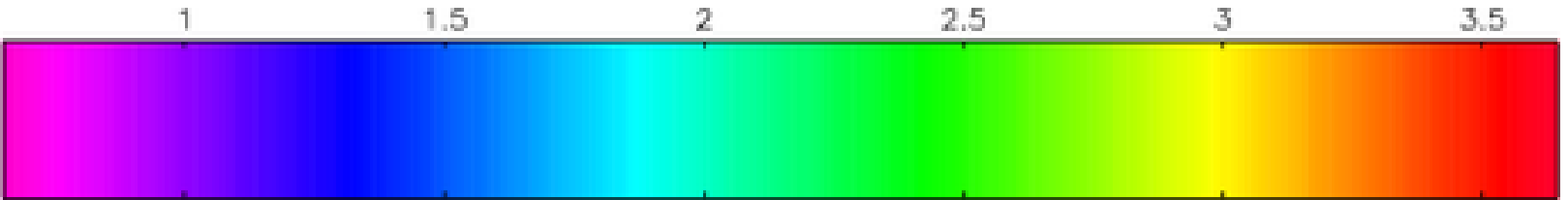}
\includegraphics[width=6.cm, height=7cm]{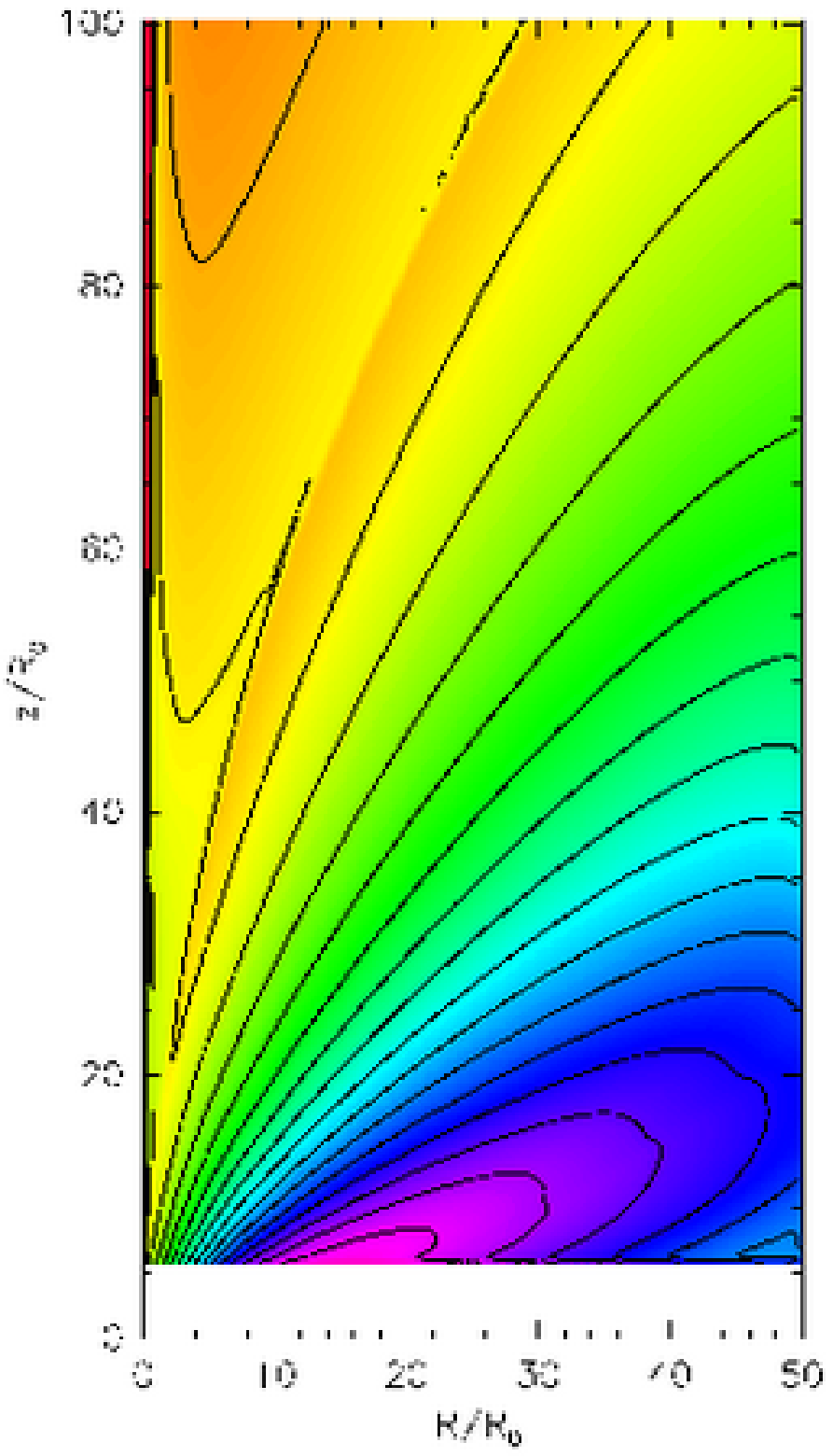}
\includegraphics[width=6.0cm,height=6.cm]{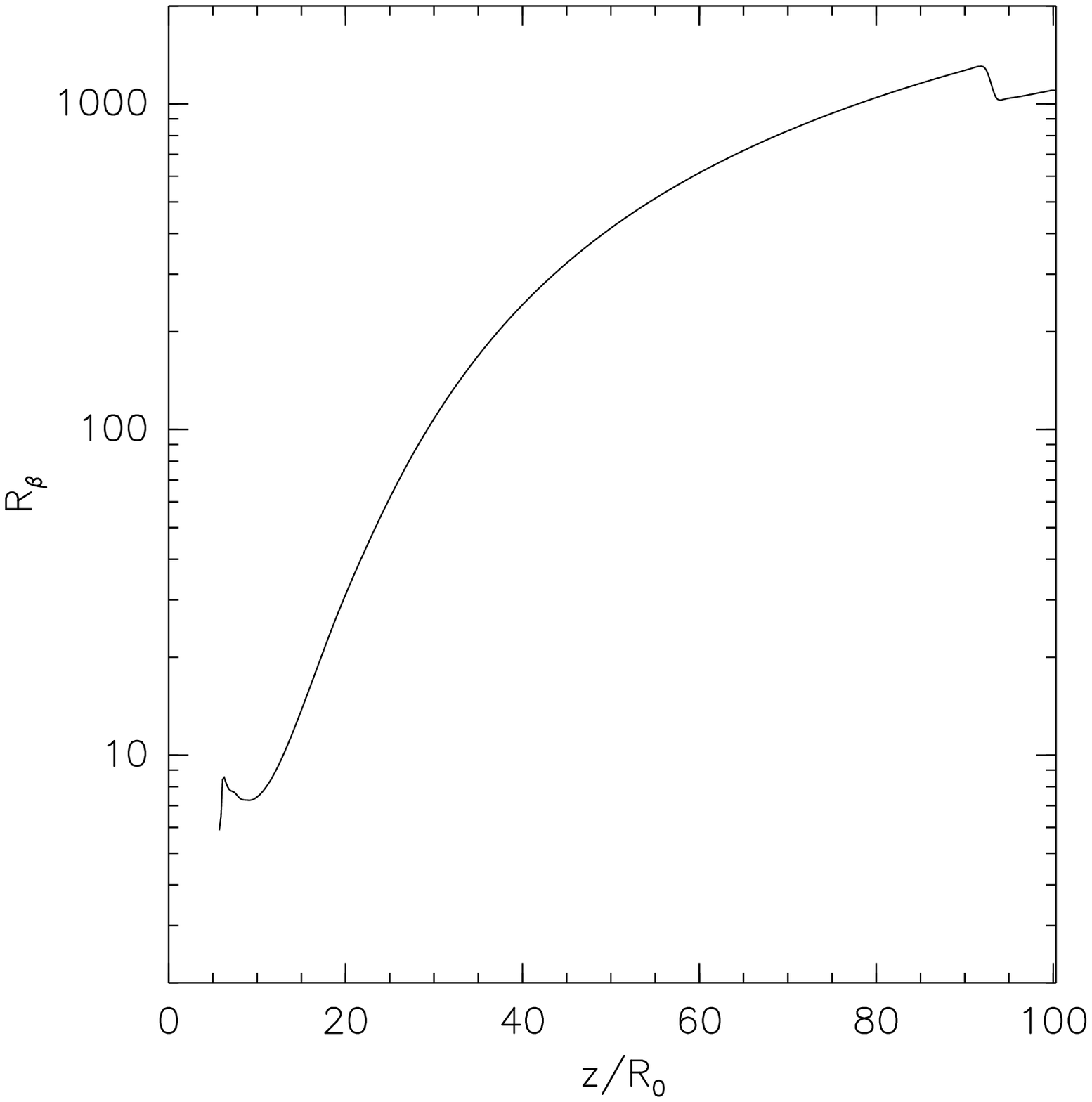}
\caption{Values of $\rbeta$ for the final state with $\etahat=0.03$, at resolution
$R\times Z=256\times 512$. The {\em top} panel shows contours of $\log_{10}\rbeta$ (six
contour lines per decade).  It is also shown in colour grading. The {\em bottom} panel shows values of
$\rbeta$ along the vertical line indicated in the top panel. We see that
the value just above the disk is $\rbeta=8.0$ in the slice taken here.
In the whole domain, the minimum value is $\rbeta=4$.
}
\label{rbet}
\end{figure}

Fig.~\ref{rmags2} shows the behaviour of $R_{\rm m}$ during the simulation with
the low magnetic diffusivity $\etahat=0.03$, and Fig.~\ref{rbet} shows
$\rbeta$. In the captions of these figures listed are also the values of
the $R_{\rm m}$ and $\rbeta$ at the bottom of the flow.

The corresponding minimum values for the simulation 
with super-critical $\eta$ shown in Fig.~\ref{eta1a} are 
$R_{\rm m}=5$ and $\rbeta=0.1$ at the bottom of the flow.
These values, however, are not constant in time 
since this solution is not stationary, with the dense "wing" sweeping
the computational box quasi-periodically.

Astrophysical jets, which are presumably launched from the accretion disk around
a central object, are present in various scales and around objects of various masses.
The magnetic resistivity of the disk - an essential part of the accretion mechanism in
some models - would be transported to the disk corona immediately above
the disk. 
Therefore, the effects investigated in this paper can be related to
astrophysical objects. We concentrate on the case of jets associated with 
young stellar objects,
but since our numerical simulations are scalable to objects of any mass, similar
scaling to the jets around e.g. a black hole, is possible (as long as the flow
remains non-relativistic).

Let us now scale our simulated solutions to the case of jets associated with young stellar objects.
This is readily done by inspecting the equations in the \S \ref{sec2}.
As we have normalised $\eta$ by
$V_0 R_0$, scaling to our physical system is straightforward.

The velocity $V_0$ is related to the mass of the central object by the first
of Eqs. (\ref{norms}). This leaves us with the expression for
$\eta$
\beq
\eta=\etahat V_0 R_0=\etahat \frac{\sqrt{{\cal GM}R_0}}{\kappa}\,.
\eeq
Therefore, for any object of mass ${\cal M}$ we define the unit radial
distance $R_0$, and we can estimate the physical $\eta$ in our simulations.

For young stellar objects, with mass ${\cal M} \approx M_\odot$ and characteristic
distance of $\approx$ 0.1AU, i.e.
$R_0=20 R_\odot =1.4 \times 10^{10}$m, in our setup,
$\eta =6.8 \etahat \times 10^{14}$ m$^2$s$^{-1}$.
This gives for the physical diffusivity, $\eta=2 \times 10^{13}$ m$^2$s$^{-1}$
for our small $\etahat$ value of 0.03, and
$\eta=10^{14}$ m$^2$s$^{-1}$ for the last subcritical value of $\etahat=0.15$.

\section{Summary}
In this paper we presented resistive numerical simulations of outflows with radially
self-similar initial conditions. The analytical solution of V00
has been modified, following GVT06, in a way that it provided a consistent setup
for simulations which aimed to extend the failing analytical solution in the close
vicinity of the axis, and large distances from the disk. These ideal MHD solutions
have been confirmed also by Matsakos et al. (2008), by using the PLUTO code, which
is using different numerical methods than the NIRVANA code, used in our simulations.

From the outset, it is not obvious at all that the resistive MHD solutions for such a problem should
stay close to the ideal MHD solutions. However, we find that the MHD solution changes smoothly,
with a continuous trend for the physical variables, as the resistivity increases.
Resistive solutions also reach a well defined stationary state. This in itself is
already an interesting and valuable result of the present study. The topology
of the moderately resistive MHD solutions turns out to be similar to the ideal MHD
case.

This is the case until some critical magnetic diffusivity is reached,
when the solutions become increasingly nonconservative in energies and fluxes.
The critical transition can be measured through a new dimensionless quantity $\rbeta$,
which measures the influence of resistive effects in the energy equation.

Energy and flux considerations in the present paper are of
unprecedented exactness, when it goes for stationarity of the compared runs, 
which can help to reach some conclusions on resistive MHD behaviour of
jets in the astrophysical context. Especially it could be of value for the treatment
of the disk corona nearby the disk, where the resistivity of the disk is probably
transported up to some height above the disk. The result that the solutions follow
the stable {\em trends}, confirms intuitive expectation.

However, the existence of the critical magnetic diffusivity, now
illustrated clearly for the first time in comparison with the analytical
solution of the closely related ideal-MHD problem, sets a limit for such intuitive
reasoning.

A general conclusion for the resistive MHD simulations of jets is that they are
similar to the ideal MHD
solutions, for a finite range of the parameter of magnetic diffusivity. In this 
range, they reach a well defined stationary state. This also
extends to the
{\em numerical resistivity} implicitly present in codes. In this respect
our result confirms that in numerical simulations with reasonable resolution,
the result should not differ significantly from the ideal MHD solution. Departure
of the solution from the ideal-MHD regime seems to occur, at least for simple,
smooth initial and boundary conditions, only for larger values of the magnetic
diffusivity, a few orders of magnitude above the level of numerical magnetic 
diffusivity. This regime of our solutions will be investigated in more detail in a following study.

Here we presented an application of our results to the case of jets associated with 
young stellar objects. Evidently, these results are scalable to various other 
astrophysical cases, as well.

\section*{Acknowledgments}
The present work was supported
by the European Community's Marie
Curie Actions - Human Resource and Mobility within the JETSET (Jet
Simulations, Experiments and Theory) network under contract MRTN-CT-2004
005592, in Athens. M\v{C} finished this work in TIARA, and expresses gratitude
 to TIARA/ASIAA in Taiwan for the possibility to use their Linux clusters
for performing the final simulations. @JG for the first ten. The authors 
would like to thank an anonymous referee for pointing out relevant references and
helpful comments.

\bsp

\label{lastpage}

\end{document}